\documentclass[11pt, a4paper, oneside, reqno]{amsart}
\ifdefined\pdfminorversion
\fi

\usepackage[utf8]{inputenc}
\usepackage[T1]{fontenc}
\usepackage{xcolor}
\usepackage{bm}
\usepackage{booktabs}
\usepackage{microtype}
\usepackage{placeins}
\usepackage{graphicx}
\usepackage{caption}
\usepackage{subcaption}
\usepackage{float}
\usepackage{comment}
\usepackage[norelsize,ruled,vlined]{algorithm2e}
\usepackage{listings}
\usepackage{pgfplots}
\pgfplotsset{compat=1.18}

\definecolor{darkblue}{rgb}{0.0, 0.0, 0.45}
\definecolor{lightblue}{rgb}{0.94, 0.97, 1.0}  
\definecolor{lightblue2}{rgb}{0.68, 0.85, 0.9}
\definecolor{lightcyan}{rgb}{0.88, 1.0, 1.0}
\definecolor{palepink}{rgb}{0.98, 0.85, 0.87}
\definecolor{Mahogany}{rgb}{0.75, 0.25, 0.0} 
\definecolor{ForestGreen}{rgb}{0.13, 0.55, 0.13}  

\allowdisplaybreaks
\date{\today}
\makeatletter
\def\@settitle{\begin{center}%
		\baselineskip14\p@\relax
		\normalfont\LARGE\scshape\bfseries
		\@title
	\end{center}%
}

\def\@setauthors{%
  \begingroup
  \def\thanks{\protect\thanks@warning}%
  \trivlist
  \centering\footnotesize \@topsep30\p@\relax
  \advance\@topsep by -\baselineskip
  \item\relax
  \author@andify\authors
  \def\\{\protect\linebreak}%
  \authors%
  \ifx\@empty\contribs
  \else
    ,\penalty-3 \space \@setcontribs
    \@closetoccontribs
  \fi
  \endtrivlist
  \endgroup
}

\makeatother
\makeatletter

\def\subsection{\@startsection{subsection}{2}%
	\z@{.5\linespacing\@plus.7\linespacing}{.5\linespacing}%
	{\normalfont\large\bfseries}}

\def\subsubsection{\@startsection{subsubsection}{3}%
	\z@{.5\linespacing\@plus.7\linespacing}{.5\linespacing}%
	{\normalfont\itshape}}

\usepackage[
  colorlinks=true,
  raiselinks=true,
  linkcolor=darkblue,
  citecolor=Mahogany,
  urlcolor=ForestGreen,
  pdfauthor={Mohsen Sadr},
  pdftitle={},
  pdfkeywords={},
  pdfsubject={},
  plainpages=false
]{hyperref}

\newcommand{\orbfortran}{ORB5}
\newcommand{\orbx}{ORB5X}
\newcommand{\performancechartheight}{0.3\linewidth}

\title[\orbx{}]{\orbx{}: a performance-portable global electromagnetic gyrokinetic PIC code in C++/Kokkos built using Agentic AI}
\author{Mohsen Sadr\textsuperscript{\dag}}
\thanks{\textsuperscript{\dag}Corresponding author.}
\thanks{M. Sadr: Paul Scherrer Institute, CH-5232, Villigen, Switzerland, 
and Department of Mechanical Engineering, MIT, Cambridge, MA 02139, USA.
Emails: \href{mailto:mohsen.sadr@psi.ch}{\texttt{mohsen.sadr@psi.ch}}
and \href{mailto:msadr@mit.edu}{\texttt{msadr@mit.edu}}.}

\author{Emmanuel Lanti}
\thanks{E. Lanti: \'Ecole Polytechnique F\'ed\'erale de Lausanne (EPFL), SCITAS, CH-1015 Lausanne, Switzerland.}

\author{Alexey Mishchenko}
\thanks{A. Mishchenko: Max-Planck-Institut f\"ur Plasmaphysik, D-17491 Greifswald, Germany.}

\author{Xin Wang}
\thanks{X. Wang: Max-Planck-Institut f\"ur Plasmaphysik, D-85748 Garching, Germany.}

\author{Laurent Villard}
\thanks{L. Villard: \'Ecole Polytechnique F\'ed\'erale de Lausanne (EPFL), Swiss Plasma Center (SPC), CH-1015 Lausanne, Switzerland.}

\SetKwInput{KwInput}{Input}
\SetKwInput{KwOutput}{Output}

\definecolor{codegreen}{rgb}{0,0.6,0}
\definecolor{codegray}{rgb}{0.5,0.5,0.5}
\definecolor{codepurple}{rgb}{0.58,0,0.82}
\definecolor{backcolour}{rgb}{0.95,0.95,0.92}

\lstdefinestyle{mystyle}{
    backgroundcolor=\color{backcolour},
    commentstyle=\color{codegreen},
    keywordstyle=\color{magenta},
    numberstyle=\tiny\color{codegray},
    stringstyle=\color{codepurple},
    basicstyle=\ttfamily\footnotesize,
    breakatwhitespace=false,
    breaklines=true,
    captionpos=b,
    keepspaces=true,
    numbers=left,
    numbersep=5pt,
    showspaces=false,
    showstringspaces=false,
    showtabs=false,
    tabsize=2
}
\begin{document}

\begin{abstract}
\orbx{} is a C++17/Kokkos AI-assisted translation of \orbfortran{},
 a global electromagnetic gyrokinetic particle-in-cell code for toroidal confined plasmas.
 The translation preserves the mathematical model and
the main numerical algorithms of the Fortran implementation, 
i.e. a Lagrangian gyrokinetic formulation, 
marker-particle representation of the distribution functions, 
finite-element B-spline representation of the fields, 
Fourier filtering in the angular directions, 
MPI domain decomposition and cloning, 
and HDF5 diagnostic output. 
The main software change is the replacement of Fortran modules and allocatable arrays by typed C++ classes, structs, namespaces, 
and Kokkos Views and kernels for performance portability across multicore CPU and GPU backends. 
This paper summarizes the physical models, 
the numerical scheme, 
the distributed and shared-memory parallel design, the AI-assisted code translation workflow,
and the validation path used to compare \orbx{} against the original Fortran reference codebase. 
The initial tests for ITG, ITPA, and chirping show an agreement of almost machine accuracy 
between \orbx{} and \orbfortran{} for the electrostatic and electromagnetic field components.
Furthermore, we demonstrate that with the modern CMake and Kokkos,
\orbx{} can be easily compiled and efficiently run on personal Linux and Mac laptops, 
as well as on high-performance computing clusters such as LUMI-G, Daint-ALPS, CINECA, and Discoverer.

\end{abstract}

\maketitle

\section{Introduction}

Global gyrokinetic simulations are a central tool for studying microturbulence, 
energetic-particle physics, 
Alfv\'enic waves, 
and nonlinear transport in magnetically confined plasmas. 
\orbfortran{} is one of the established global electromagnetic gyrokinetic particle-in-cell (PIC) codes in toroidal geometry \cite{lanti2020orb5}.
 Its production implementation is written in Fortran 
 and has accumulated a broad physics model, 
 a mature input/output ecosystem, 
 and highly tuned MPI/OpenMP/OpenACC parallelism. 
The academic time constraints in the development of \orbfortran{} have marginalized  long-term maintainability and portability issues.
 Furthermore, the reliance on Fortran-support of NVIDIA compiler as well as its dependencies have hindered portability to new architectures,
 leading to significant operational cost.

\orbx{} is a source-to-source translation of \orbfortran{} into C++17 with Kokkos \cite{trott2022kokkos}.
 The goal is not to introduce a new gyrokinetic model, 
 but to preserve the Fortran behavior while making the codebase easier to compile, 
 test, 
 reason about,
 and retarget to modern CPU and GPU systems. 
 During the translation, the Fortran sources were kept side by side with their C++ counterparts. 
This allowed direct module-by-module comparison. 
For example, the new code mirrors the file naming such as \texttt{fields.F90} with \texttt{fields.cpp}.

In this work, we first record the physical and numerical model implemented by \orbx{} 
in a form that is independent of either programming language. 
Second, we document how the Fortran parallel structure was translated into an MPI plus Kokkos design. 
Third, we give the validation strategy used to establish equivalence between \orbx{} and \orbfortran{}, 
from unit tests of individual kernels to full ITG, ITPA, and Chirping field comparisons produced in HDF5 format.

\section{Physical Model}

\orbx{} advances gyrocenter marker particles in an axisymmetric tokamak equilibrium 
and solves the global electromagnetic gyrokinetic Vlasov--Maxwell system in a PIC representation. 
Here, we deploy the variational formulation of gyrokinetics with drift-kinetic or reduced electron models, 
electrostatic and electromagnetic field equations, optional spliting of pullback scheme for electromagnetic simulations,
collisions, 
noise-control operators, 
heating/source terms, 
and strong-flow extensions.
ORB5X uses the ORB5 axisymmetric tokamak-equilibrium representation, 
straight-field-line magnetic coordinates, normalized code units, 
a $\delta f$ control-variate marker formulation, 
and field unknowns given by the electrostatic potential $\phi$ and,
 in electromagnetic runs, the parallel vector potential $A_\parallel$. 
 The explicit equilibrium formulas, 
 normalizations, 
 Hamiltonians, 
 weak field equations, 
 and model variants are collected in Appendix~\ref{app:physical_model}.

The translation preserves the physical model rather than introducing a new one. 
Therefore, the most important model statements for ORB5X are implementation invariants,
i.e. the marker attribute layout must remain compatible with the original time integration, 
collision, 
diagnostic, 
and migration paths; 
the weak quasineutrality and Amp\`ere equations must assemble the same finite-element systems; 
and the electromagnetic split/pullback workflow must preserve the intended control-variate algebra. 
These invariants are used throughout the verification strategy described below.

\section{Numerical Discretization and Algorithms}

The ORB5X time step follows the standard global gyrokinetic PIC cycle. 
Marker orbits and weights are advanced with an operator-split scheme 
in which collisionless dynamics are followed by collisions, 
sources, and noise-control operators. 
Marker quantities are deposited onto spline coefficients; 
the quasineutrality and Amp\`ere systems are solved in a tensor-product B-spline/Fourier basis; 
and the resulting fields and gradients are interpolated back to marker positions. 
The main numerical ingredients are summarized here, 
while the detailed formulas, loading conventions, filters, gyroaveraging, and diagnostic balances
 are collected in Appendix~\ref{app:numerical_model}.

The field representation uses B-spline finite elements on an $(N_s,N_\chi,N_\varphi)$ grid 
and Fourier filtering in the angular directions. 
The angular field solve retains the ORB5 structure, 
i.e. poloidal transforms, distributed toroidal transposes, toroidal transforms, mode filtering, 
and inverse transforms before the finite-element solve. 
Gyroaveraging is evaluated by sampling a Larmor ring in the poloidal plane.
 Noise and profile-control machinery includes modified Krook damping, 
 coarse graining, quadtree smoothing, thermal relaxation, and fixed-power heating operators.

In C++, performance-critical particle and field arrays are represented with Kokkos Views, 
and data-parallel loops are expressed with \texttt{Kokkos::parallel\_for} 
and \texttt{Kokkos::parallel\_reduce} where the translating has exposed regular loop structure. 
Some routines still stage through host mirrors because FFTW, 
HDF5, LAPACK, or MPI communication expects host-accessible buffers. 
These host-staged paths are part of the numerical implementation
 and are reported separately from fully device-resident kernels in the portability discussion.

\section{Parallelism and HPC Design}

\begin{figure}[htbp]
\centering
\includegraphics[width=0.72\linewidth]{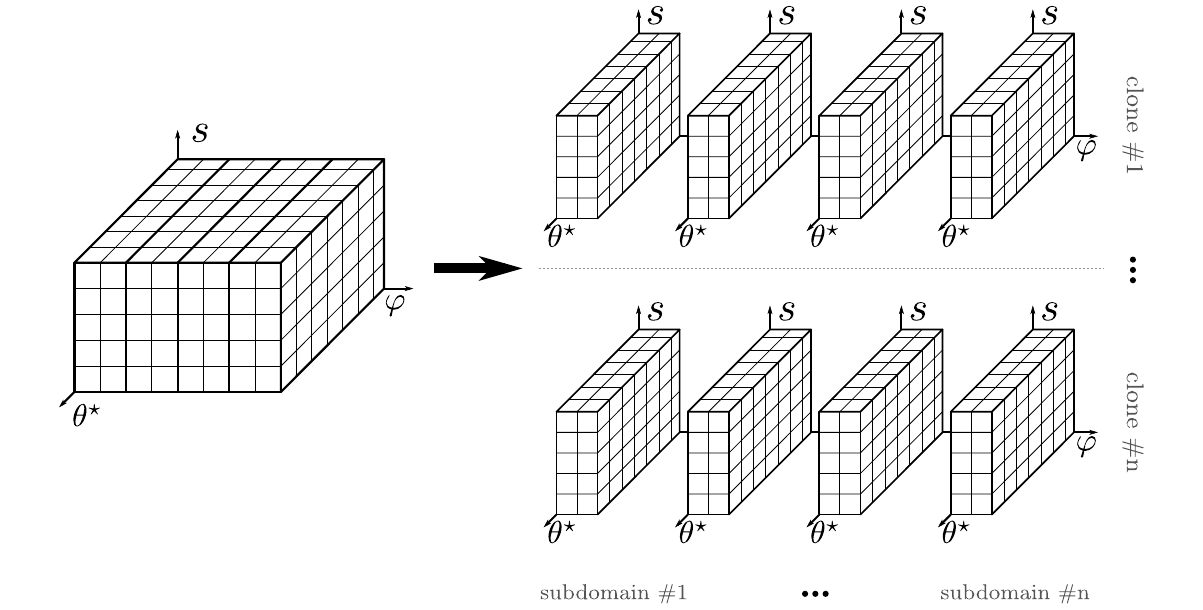}
\caption{MPI parallelization using toroidal domain decomposition and domain cloning \cite{lanti2020orb5}.}
\label{fig:orb5x_parallel_topology}
\end{figure}

\orbfortran{} uses toroidal domain decomposition as its spatial MPI decomposition 
and combines it with domain cloning of marker populations \cite{Kim2000,Hatzky2006,lanti2020orb5}. 
The corresponding schematic is shown in Fig.~\ref{fig:orb5x_parallel_topology}.
 If $P_{\rm sub}$ is the number of toroidal subdomains 
 and $P_{\rm clones}$ is the number of clones, 
 then the total number of MPI ranks is
\begin{equation}
  P_{\rm MPI}=P_{\rm sub}P_{\rm clones}.
\end{equation}
The physical grid is replicated across clones, 
while markers are equally divided among the clones.
Within each clone, the physical domain is split only in the toroidal direction, 
giving a periodic one-dimensional subdomain communicator. 
Thus clone communication is dominated by reductions of deposited grid quantities, 
whereas subdomain communication contains nearest-neighbor guard-cell exchange, 
global transposes of field data for Fourier transforms,
 and point-to-point particle migration. 
 The original \orbfortran{} paper emphasizes that particle exchange is not restricted to nearest neighbors, 
 i.e. particles may cross multiple toroidal subdomains during one step, 
 and the exchange algorithm supports all-to-all communication within a clone. 
 The same section of the ORB5 paper also notes the performance balance, 
 i.e. toroidal domain decomposition scales well, 
 while a large number of clones increases replicated field memory and reduction cost, 
 so shared-memory parallelism is used to reduce the clone count required for a given run.


\orbx{} preserves this topology explicitly.  The C++ global state mirrors the Fortran \texttt{parallel} module through  a \texttt{ParallelVariables} object containing communication variables.   The routine \texttt{Pputil::pptopology} is the C++ counterpart of \texttt{pptopology}, i.e. it creates a two-dimensional Cartesian communicator,  makes the subdomain dimension periodic, splits ranks with the same clone index and the ranks with the same toroidal-subdomain index. This design intentionally keeps the MPI ownership model unchanged, so numerical differences between \orbx{} and \orbfortran{}   can be audited inside the kernels rather than being mixed with a new decomposition.

Particle data are owned by the MPI rank corresponding to the marker's clone and toroidal subdomain. 
The C++ particle containers keep the Fortran species/module organization 
but store performance-critical arrays as Kokkos Views, 
including marker phase-space coordinates, weights, flags, gyropoint work arrays, and diagnostic accumulators. 
The dominant particle loops
  are expressed with \texttt{Kokkos::parallel\_for} and \texttt{Kokkos::parallel\_reduce}. 
  This replaces the OpenMP/OpenACC loop directives used in \orbfortran{} 
  with an execution-space abstraction selected at the build time.

Particle migration
in the C++ module follows the Fortran 
algorithm 
but makes the datatype handling explicit with C++ templates. 
A Kokkos kernel first counts particles that have left the local toroidal slab.
 The ranks then exchange counts with \texttt{MPI\_Ialltoall}, 
 build send and receive displacements,
  pack the leaving marker attributes into typed buffers,
   post nonblocking receives, 
   send the outgoing buffers, 
   and unpack received particles into either holes left by departing particles 
   or appended slots at the end of the local arrays. 
   Remaining holes are filled by backfilling from the tail of the particle arrays. 

\sloppy Field data follow a different communication pattern. 
Deposited quantities such as charge density, current, and moments are first accumulated locally 
from the markers on each clone.
ORB5X then reduces clone-local contributions, 
 applies subdomain guard updates,
 and performs the angular Fourier path.
 Similarly, data in ORB5X are also transformed in the poloidal angle, 
 transposed across toroidal subdomains using \texttt{MPI\_Sendrecv}, 
 transformed and filtered in the toroidal angle, 
 and transposed back before the finite-element field solve and interpolation. 
 ORB5X also contains the local-DFT routines 
 which uses partial toroidal transforms and an MPI \texttt{reduce\_scatter}-style reduction to reduce the need for full transposes.

The matrix-solver interface also retains the original communication structure. 
\orbx{} creates the assembly communicator, 
distributes finite-element matrix assembly work,
 and broadcasts assembled matrices to the clone group.
 The solver therefore sees the same mathematical systems as in \orbfortran{}, 
 while the surrounding C++ layer makes communicator ownership and data movement explicit.

Runtime ordering is an important part of the ORB5X design. 
The executable initializes MPI first, 
initializes Kokkos before any global Kokkos Views are constructed, 
allocates module state,
runs the control layer, 
and then releases global state before \texttt{Kokkos::finalize}. 
Parallel HDF5 files are closed while MPI is still active, 
because the HDF5 MPI-IO layer owns duplicated communicators. 
The current executable then terminates without running late static destructors, 
avoiding destruction of static Kokkos Views after the Kokkos runtime has shut down. 
This is a C++ runtime-management detail absent from the Fortran code, 
but it is essential for robust MPI/Kokkos/HDF5 execution.

Shared-memory and device parallelism are therefore layered below the preserved MPI algorithm. 
\orbfortran{} used OpenMP and
 OpenACC directives for particle-loop acceleration. 
 ORB5X instead uses Kokkos Views and kernels, so the same source can target
  serial, threads, OpenMP, CUDA, HIP, or other Kokkos-supported backends when the required libraries are available.
  The principal portability boundary is not the particle kernels themselves, 
  but the parts of the algorithm that still stage through host-accessible buffers for MPI, FFTW, LAPACK, and HDF5.

\section{Translation and Verification Workflow}

The translation was performed with the Fortran and C++ sources available in the same tree. 
This made it possible to preserve names, 
constants, and array semantics where they encode numerical assumptions. 
Many C++ comments explicitly cite the corresponding Fortran file and line region. 
Examples include the field allocation bounds, 
B-spline interpolation stencils,
 Fourier filter construction, 
 solver-matrix indexing, gyropoint formulas, reduced-weight algebra, and diagnostic output layout.

\begin{figure}[H]
\centering
\includegraphics[width=0.98\linewidth]{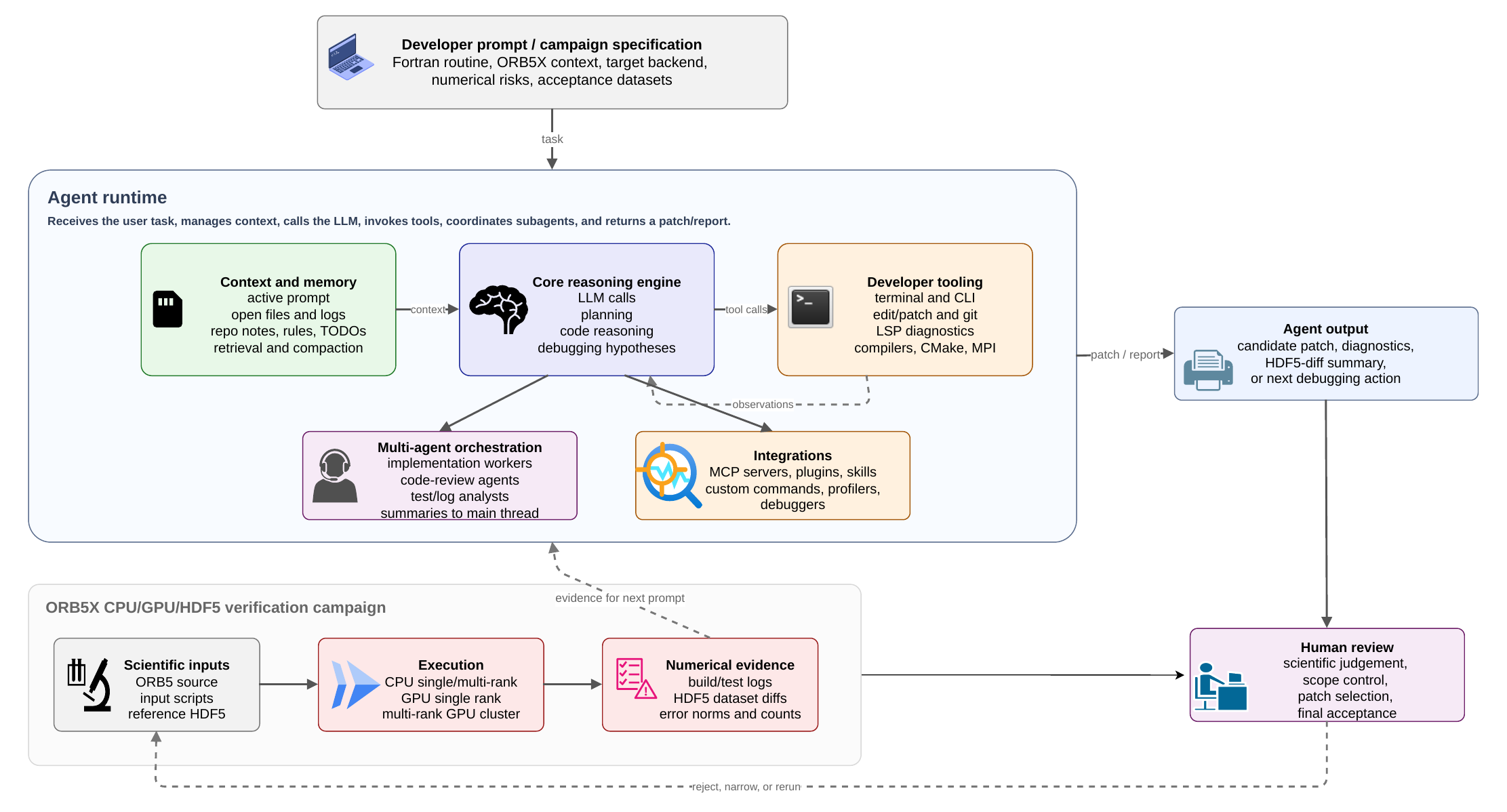}
\caption{Agentic-AI workflow used in the ORB5X translation. The structure reflects current coding-agent practice, i.e. a core LLM thread is supplied with compacted project context, memory, specialized agents, permissioned developer tools, language-server and compiler feedback, and automated execution results. In ORB5X, acceptance remained tied to developer review and numerical evidence, especially HDF5 parity against the Fortran reference across CPU, GPU, and multi-rank configurations.}
\label{fig:agentic_ai_workflow}
\end{figure}

The source translation was assisted by an agentic-AI workflow using several coding agents 
and LLM-assisted development environments, including Cursor, Claude Code, Open Code, and Codex
equipped with local LLMs such as \texttt{qwen2.5-coder:32b} and \texttt{gpt-oss:120b} served by Ollama on the local cluster.
These systems were used as implementation 
and debugging assistants rather than as unchecked translators.
A typical campaign began with a narrowly scoped prompt containing the Fortran routine,
 the corresponding ORB5X module context,
  array bounds, relevant derived types, and the benchmark or HDF5 datasets used for acceptance.
   The agent then produced a candidate C++ implementation, 
   an array-mapping note, and a list of numerical risks such as lower-bound offsets,
    reductions, MPI collectives, or host/device copies. 
    The human developer reviewed the patch, selected the useful parts, and launched the next verification campaign
    Fig.~\ref{fig:agentic_ai_workflow}.

Each implementation campaign used the same acceptance principle, 
i.e. a change was kept only if it preserved or improved numerical agreement with the Fortran reference. 
The first gate was a CPU single-rank run, 
which isolates local algorithmic issues such as indexing, loop bounds, spline stencils, and floating-point update order.
 The second gate was a CPU multi-rank run, which adds MPI topology, guard-cell exchange, reductions over clones,
  particle migration, transposes, and HDF5 collective output. 
  Once the CPU path was stable, the same benchmark was run on a GPU backend to expose Kokkos memory-space mistakes,
   missing deep copies, race conditions in kernels, 
   and host-staged library calls.
    Finally, the benchmark was repeated on multiple GPU ranks on the cluster, 
    where both device execution and distributed-memory semantics had to agree with the reference output.

The role of the agentic workflow was most useful in the debugging loops between these gates.
 When an HDF5 comparison showed a new discrepancy,
  the agent was given the changed datasets, norms, rank count, backend, 
  recent patch, and the relevant source routines.
   It was then asked to propose a small debugging hypothesis and a minimal patch or diagnostic. 
   This produced an iterative process: implement a bounded change, build, 
   run ORB5X and the Fortran reference when needed, compare HDF5 files dataset-by-dataset,
    and accept the change only when the discrepancy count or error norm decreased,
     or when the discrepancy remained unchanged while enabling the next required porting step. 
     This kept the use of LLMs tied to reproducible numerical evidence rather than subjective code similarity.

Verification was organized in layers. 
The lowest layer consists of GoogleTest unit tests for numerical kernels and data structures.
 The current CMake test suite registers tests for spline bases, Fourier filters, 
 boundary conditions, solver data, solver MPI formulas, gyropoints, 
 mass matrices, particle-field interpolation, random generators, collisions,
  quadtree noise control, sources, background profiles, diagnostics, HDF5 output,
   and orchestration surfaces. 
   Several tests are explicitly written as Fortran-parity tests 
   and encode formulas or edge cases from the original source. The second layer compares output files and diagnostic fields from complete runs.
 For the ITG, ITPA, and Chirping benchmark, 
 the local repository contains ORB5X and reference HDF5 output files together with plotting
  and comparison scripts. 

\section{Code Design}

The initial ORB5X design keeps the physical module boundaries of \orbfortran{} 
while making the dependencies explicit in C++. 
The top-level library \texttt{orb5x\_lib} 
is built from C++ classes and module-like components for particles, fields, sources, 
collisions, diagnostics, profiles, equilibrium,
parallel particle movement, and solver infrastructure. 
Here ``module-like'' refers to the organization of the code into physics and numerical components, 
not to C++20 language modules.
The executable \texttt{orb5x} initializes MPI and Kokkos, initializes global code state, 
calls the control layer, and performs ordered runtime cleanup.

The C++ design is object-oriented primarily through class encapsulation, composition, and 
typed state structs
rather than through an inheritance hierarchy. 
For example, \texttt{InterfaceSolverInOrb5} owns the solver-interface state behind a class interface, 
deletes copy construction and copy assignment, and exposes static facade methods that preserve the original 
Fortran-module call pattern. 
Similarly, the particle, field, diagnostic, and communication paths group related state and methods into classes, 
while shared input-derived state is represented by typed structs.

An important part of this consolidation is the treatment of support libraries that were external to, 
or loosely coupled with, the original Fortran source. 
ORB5 depends on SPClibs\footnote{\url{https://gitlab.com/spclibs}} for spline and matrix infrastructure 
and on \texttt{futils}, which is a wrapper around HDF5 providing helper functions for file operations.
 In ORB5X, the numerical SPClibs functionality used by the field and solver paths 
 is refactored directly.
 It defines one-, two-, and three-direction tensor-product spline objects, 
 knot storage, Gauss quadrature data, basis-function evaluation,
  grid evaluation, coefficient transforms, and mass-matrix construction. 
  Also, the \texttt{matrix} component provides real and complex dense, banded, packed-banded, 
  and periodic matrix containers, together with update, access, factorization, backsolve, 
  matrix-vector product, determinant, and copy operations. 
  The low-level \texttt{pppack} component preserves the de Boor spline kernels used by SPClibs,
   including B-spline basis and derivative evaluation, 
   interval lookup, Gauss-Legendre quadrature, and piecewise-polynomial conversion. 
   The Fortran \texttt{futils} role is not kept as a separate dependency;
    its file creation, dataset writing, and diagnostic-output patterns are represented in the C++ diagnostics 
    and control/output layers using the ORB5X HDF5 interface.

This design gives three practical benefits. 
First, individual kernels can be compiled and tested without running a full plasma simulation. 
Second, Kokkos Views make memory spaces and deep copies visible in the source, 
which is essential for accelerator portability. 
Third, the side-by-side Fortran/C++ layout supports incremental numerical audits, 
i.e. when a parity test fails, the relevant C++ routine can be compared directly to the Fortran routine 
that originally defined the behavior.

The translation was intentionally conservative in the numerical layer 
but more opinionated in the software layer. 
Table~\ref{tab:fortran_cpp_design} summarizes the main differences between the original Fortran implementation and ORB5X, 
together with the reason for each C++ design choices. 
These are implementation differences rather than physics-model differences, 
i.e. the goal was to preserve the ORB5 algorithms 
while making ownership, memory spaces, dependencies, and tests explicit. Here, we attempted to keep the \orbx{} design as close as possible to \orbfortran{} to simplify verifications as the first implementation, knowing that the design of the original Fortran code can be greatly improved in the C++ version using Standard Template Library (STL) on the host when possible.

\begin{table}
\centering
\footnotesize
\setlength{\tabcolsep}{3pt}
\begin{tabular}{p{0.18\linewidth}p{0.32\linewidth}p{0.4\linewidth}}
\toprule
Aspect & ORB5 Fortran design & ORB5X C++ design and rationale \\
\midrule
Language and code organization & Fortran modules with global allocatable arrays. & C++17 classes, structs, namespaces, and an \texttt{orb5x\_lib} library. The physics module boundaries are kept as module-like components, but kernels can be linked into tests. \\
\addlinespace
Array ownership & Fortran lower bounds, column-major arrays, and module state. & Kokkos Views plus explicit offset mappings where ORB5 indexing semantics matter. This preserves stencils and exposes memory spaces. \\
\addlinespace
Loop parallelism & OpenMP/OpenACC directives around critical loops. & Kokkos kernels and reductions. This gives one source path for CPU threads and accelerator backends. \\
\addlinespace
MPI topology & Domain decomposition and cloning in the Fortran \texttt{parallel} module. & The same topology is kept in \texttt{ParallelVariables} and \texttt{Pputil::pptopology}. Communication changes are separated from numerical-kernel audits. \\
\addlinespace
Support libraries & External SPClibs and \texttt{futils} helper routines provide spline, matrix, and HDF5 utilities. & The needed SPClibs pieces are internal C++ components in \texttt{src/spclibs}: \texttt{bsplines}, \texttt{matrix}, and \texttt{pppack}. HDF5 helper behavior is integrated into diagnostics/control output. This makes numerical support code testable and removes hidden build assumptions. \\
\addlinespace
Particle migration & Type-specific buffers and include-file communication logic. & Templated typed paths in \texttt{parmove.cpp}. The ORB5 all-to-all migration semantics are preserved while datatype mapping is localized. \\
\addlinespace
Dependencies & Machine-specific build assumptions. & CMake can find or build Kokkos, HDF5, FFTW3, and LAPACK/OpenBLAS. This lowers the cost of laptop and new-cluster builds. \\
\addlinespace
Runtime cleanup & Finalization follows the Fortran program lifetime. & \texttt{main.cpp} explicitly orders MPI, Kokkos, global Views, HDF5, and MPI finalization to avoid late C++ static destruction problems. \\
\addlinespace
Verification & Mostly full-code benchmark validation. & Unit and parity tests link against \texttt{orb5x\_lib}; HDF5 comparisons remain the end-to-end check. \\
\bottomrule
\end{tabular}
\caption{Important implementation differences between the original Fortran ORB5 code and ORB5X. The C++ choices were made to preserve numerical behavior while improving testability, explicit ownership, dependency management, and performance portability.}
\label{tab:fortran_cpp_design}
\end{table}

\section{Portability}

ORB5X is configured with CMake and a small set of user-facing options.
 The minimum external requirements are a C++17 compiler, CMake, and an MPI implementation.
 The remaining numerical libraries, such as Kokkos, HDF5, FFTW3, and BLAS/LAPACK, can either be supplied by the system
 or downloaded and built by ORB5Xs' CMake. 
 Please note that the installed HDF5 must be Parallel HDF5 with C++ bindings for the current output layer. 
 Also note that when Kokkos is configured with OpenMP, ORB5X links OpenMP for host execution; 
 when Kokkos is configured with CUDA, the build adds the relaxed-constexpr option required
  by CUDA compilation of Kokkos Views with complex-valued data.
  The practical consequence is that a developer does not need a full production ORB5 software stack to compile ORB5X.
  See Fig.~\ref{fig:cmake_build_examples} for examples of typical build modes.

\begin{figure}[H]
\centering
\begin{minipage}{0.94\linewidth}
\tiny
\textbf{CPU/MPI configuration with dependencies built by ORB5X}
\begin{lstlisting}[language=bash,basicstyle=\ttfamily\footnotesize,breaklines=true,columns=fullflexible,frame=single,aboveskip=3pt,belowskip=6pt]
cmake -B <build dir> \
  -DORB5X_BUILD_KOKKOS_FROM_SOURCE=ON \
  -DORB5X_BUILD_HDF5_FROM_SOURCE=ON \
  -DORB5X_BUILD_FFTW3_FROM_SOURCE=ON \
  -DORB5X_BUILD_LAPACK_FROM_SOURCE=ON
cmake --build <build dir>
\end{lstlisting}
\ \\
\textbf{NVIDIA A100 GPU configuration through Kokkos/CUDA}
\begin{lstlisting}[language=bash,basicstyle=\ttfamily\footnotesize,breaklines=true,columns=fullflexible,frame=single,aboveskip=3pt,belowskip=2pt]
cmake -B <build dir> \
  -DORB5X_BUILD_KOKKOS_FROM_SOURCE=ON \
  -DKokkos_ENABLE_CUDA=ON \
  -DKokkos_ARCH_AMPERE80=ON \
  -DORB5X_BUILD_HDF5_FROM_SOURCE=ON \
  -DORB5X_BUILD_FFTW3_FROM_SOURCE=ON \
  -DORB5X_BUILD_LAPACK_FROM_SOURCE=ON
cmake --build <build dir>
\end{lstlisting}
\ \\
\textbf{AMD MI250x GPU configuration through Kokkos/HIP}
\begin{lstlisting}[language=bash,basicstyle=\ttfamily\footnotesize,breaklines=true,columns=fullflexible,frame=single,aboveskip=3pt,belowskip=2pt]
cmake -B <build dir> \
  -DORB5X_BUILD_KOKKOS_FROM_SOURCE=ON \
  -DKokkos_ENABLE_HIP=ON \
  -DKokkos_ARCH_AMD_GFX90A=ON \
  -DORB5X_BUILD_HDF5_FROM_SOURCE=ON \
  -DORB5X_BUILD_FFTW3_FROM_SOURCE=ON \
  -DORB5X_BUILD_LAPACK_FROM_SOURCE=ON
cmake --build <build dir>
\end{lstlisting}
\end{minipage}
\caption{Representative CMake configurations for CPU, NVIDIA GPU, and AMD GPU builds. The CUDA or HIP architecture flag should be changed to match the target accelerator\protect\footnotemark.
}
\label{fig:cmake_build_examples}
\end{figure}
\footnotetext{See the Kokkos configuration guide for GPU architecture options:\\ \url{https://kokkos.org/kokkos-core-wiki/get-started/configuration-guide.html}.}

\section{Verification}

In order to verify the correctness of ORB5X, 
we compare its results against established benchmarks and diagnostics from the original Fortran implementation, 
including ITG, ITPA, and Chirping test cases. Then, we report the performance and scaling for the current version of ORB5X.

\subsection{Ion Temperature Gradient (ITG)}

The ITG case is a nonlinear electrostatic ion-temperature-gradient turbulence
test based on the standard ORB5 global gyrokinetic PIC model
\cite{Jolliet2007,lanti2020orb5,donnel2021quasilinear}.  It exercises the drift-wave turbulence
workflow with kinetic deuterium ions and adiabatic electrons, so the main
physics drive is the imposed ion temperature gradient and the self-consistent
evolution of the electrostatic potential through the long-wavelength
quasineutrality equation.  In ORB5X this case is used as a heavy verification
input because it couples particle loading, mode initialization, field solve,
noise/source control, diagnostics, and MPI clone decomposition.

The input uses an ad-hoc circular equilibrium with $R_0=0.88$, $a=0.2462$,
$B_0=1.4368$, $q_0=0.854$, $q_\mathrm{edge}=3.038$, and a coefficient-based
safety-factor profile.  The fields are represented on a
$256\times512\times256$ $(s,\theta,\varphi)$ grid with quadratic B splines
and an \texttt{mn} Fourier filter retaining
$n=0,\ldots,128$ with $\Delta m=5$.  The deuterium species is kinetic with
$2.56\times10^8$ markers, fixed gyroaveraging, full nonlinear push, profile
selector \texttt{9}, gradients $\kappa_n=2.3$ and $\kappa_T=3.1$, and a
mode-seeded perturbation of amplitude $10^{-4}$ over
$m=10,\ldots,30$ and $n=10,\ldots,20$.  The electron species uses the standard
hybrid adiabatic model, 
so it does not introduce a separate electron marker population; 
it uses the same background profile parameters and an
initial perturbation over $m=12,\ldots,16$, $n=10$.  Weight-decay noise control
is enabled for the ion markers with density, zonal, and parallel-velocity
conservation, matching the ORB5 practice for long global ITG simulations
\cite{McMillan2008}.

\begin{figure}
    \centering
    \begin{tabular}{@{}c@{}}
    \includegraphics[width=\linewidth, trim=0 0 0cm 0cm, clip]{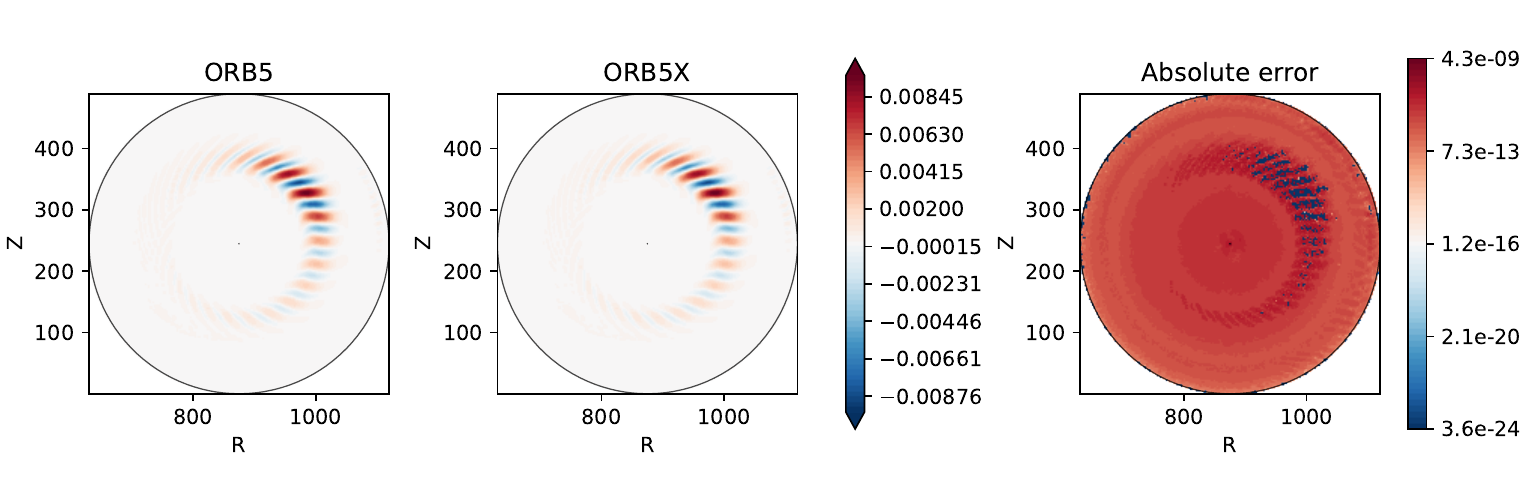}
    \end{tabular}
    \caption{ITG electrostatic-potential $\phi$ verification against the Fortran ORB5 reference. The left panel is ORB5, middle ORB5X, and right is the absolute pointwise error of $\phi$ at $t=8000\ \Omega_{c_i}^{-1}$.}
    \label{fig:itg_verification}
\end{figure}

We have simulated ITG test case using ORB5 and ORB5X, and confirmed the consistency between the output hdf5 file. As a showcase, in Fig.~\ref{fig:itg_verification} we show the consistency of the outcome electrostatic potential $\phi$ at the end of simulation using ORB5X against the original ORB5.

\FloatBarrier

\subsection{ITPA}

The ITPA case is based on the international cross-code benchmark for
fast-particle-driven toroidicity-induced Alfvén eigenmode dynamics
\cite{konies2018benchmark}.  The target mode is the $n=6$ TAE, whose dominant
poloidal harmonics are $m=-10$ and $m=-11$.  In the ORB5X input used here, this
mode is seeded directly through the \texttt{MODE\_SAW} initial perturbation
with $(n,m)=(6,-11)$ and $(6,-10)$, rather than by an external antenna \cite{sadr2022linear}.

\begin{figure}
\hspace{-2cm}
    \centering
    \begin{tabular}{ccc}
    \hspace{0.5cm} $\phi$ & $A_{||}^s$ & $A_{||}^h$
    \\
\includegraphics[width=0.35\linewidth, trim=0 0 1.4cm 2cm, clip]{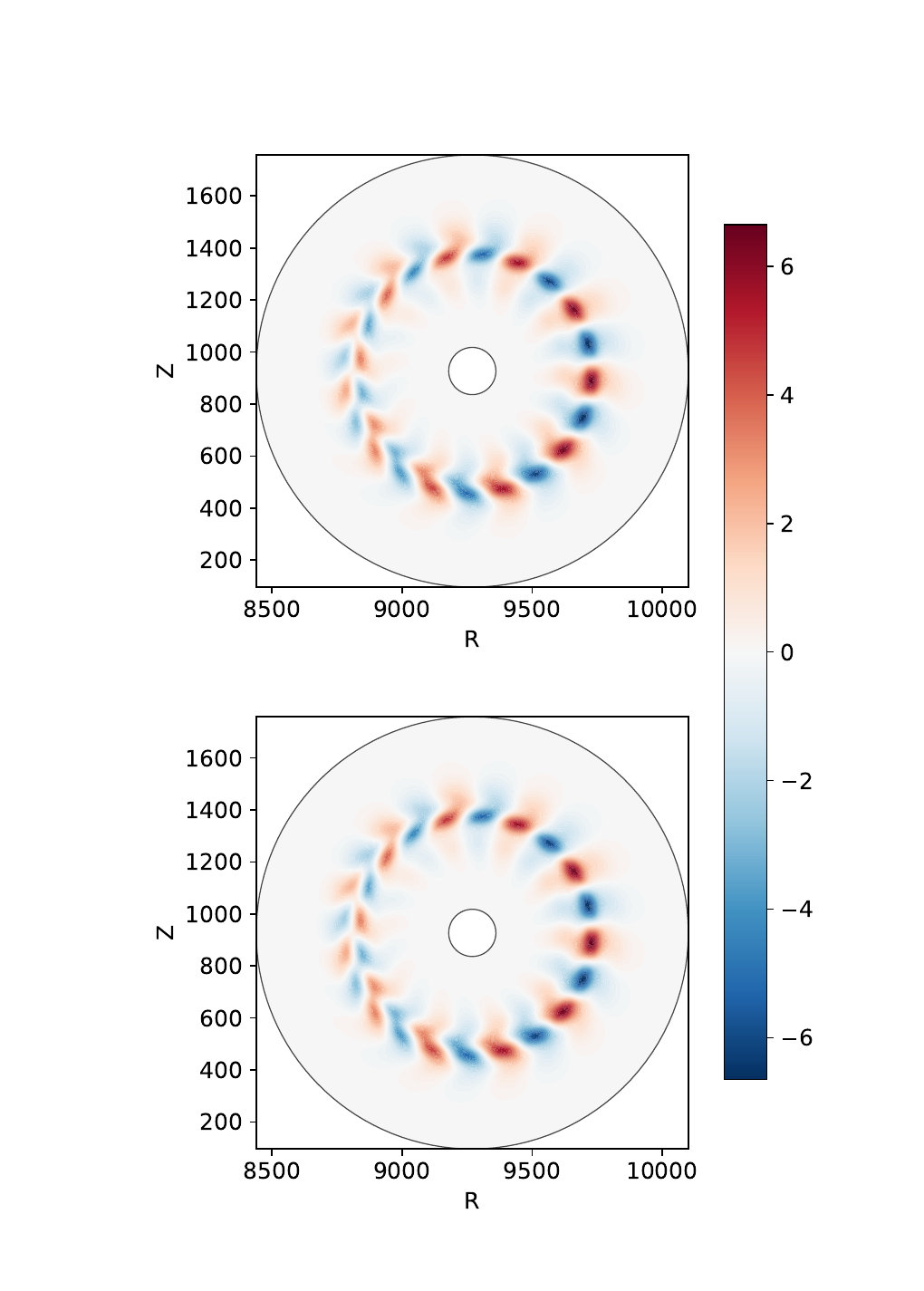}
    &
\includegraphics[width=0.35\linewidth, trim=0 0 0.9cm 2cm, clip]{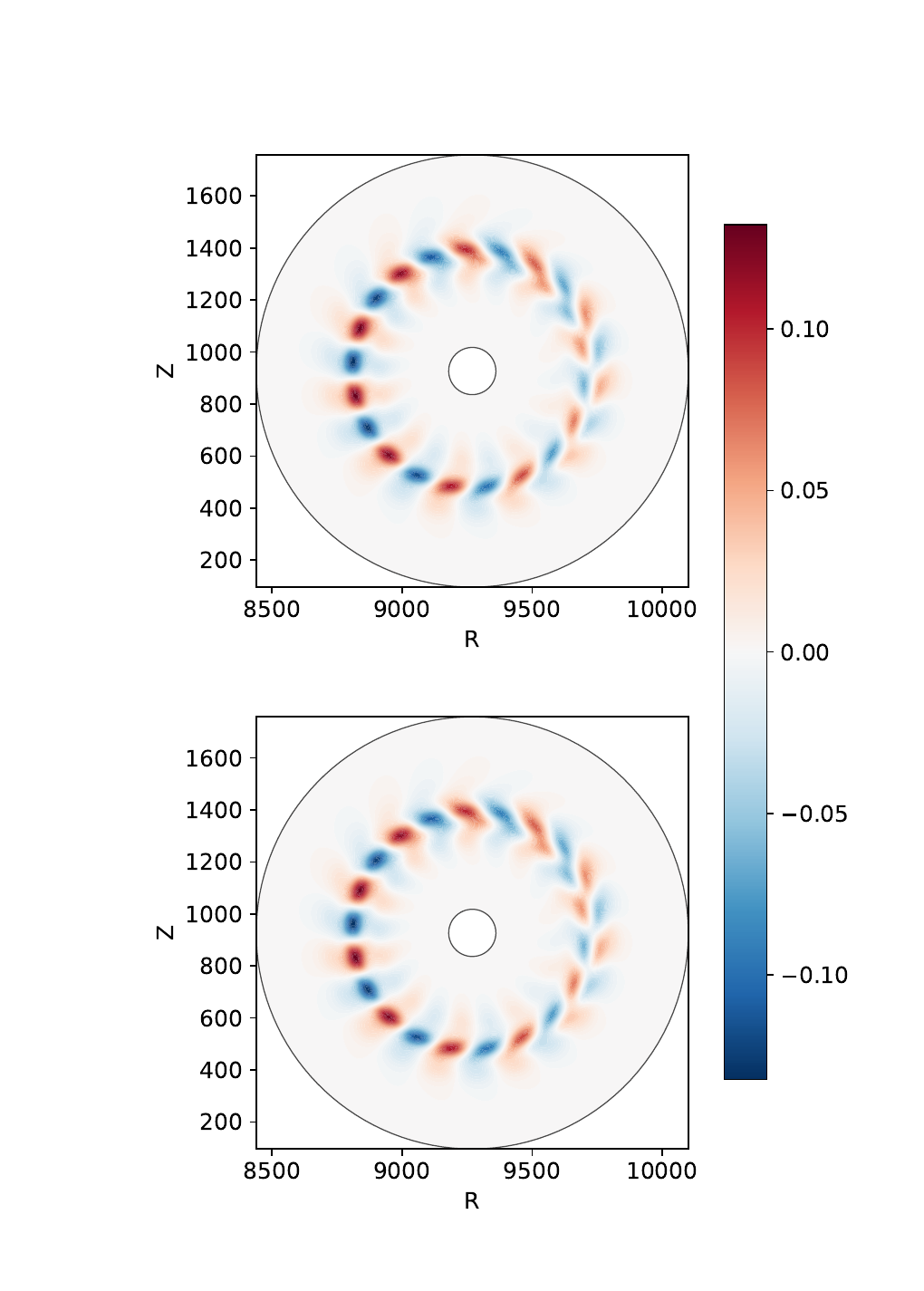}
    &
    \includegraphics[width=0.35\linewidth, trim=0 0 0.5cm 2cm, clip]{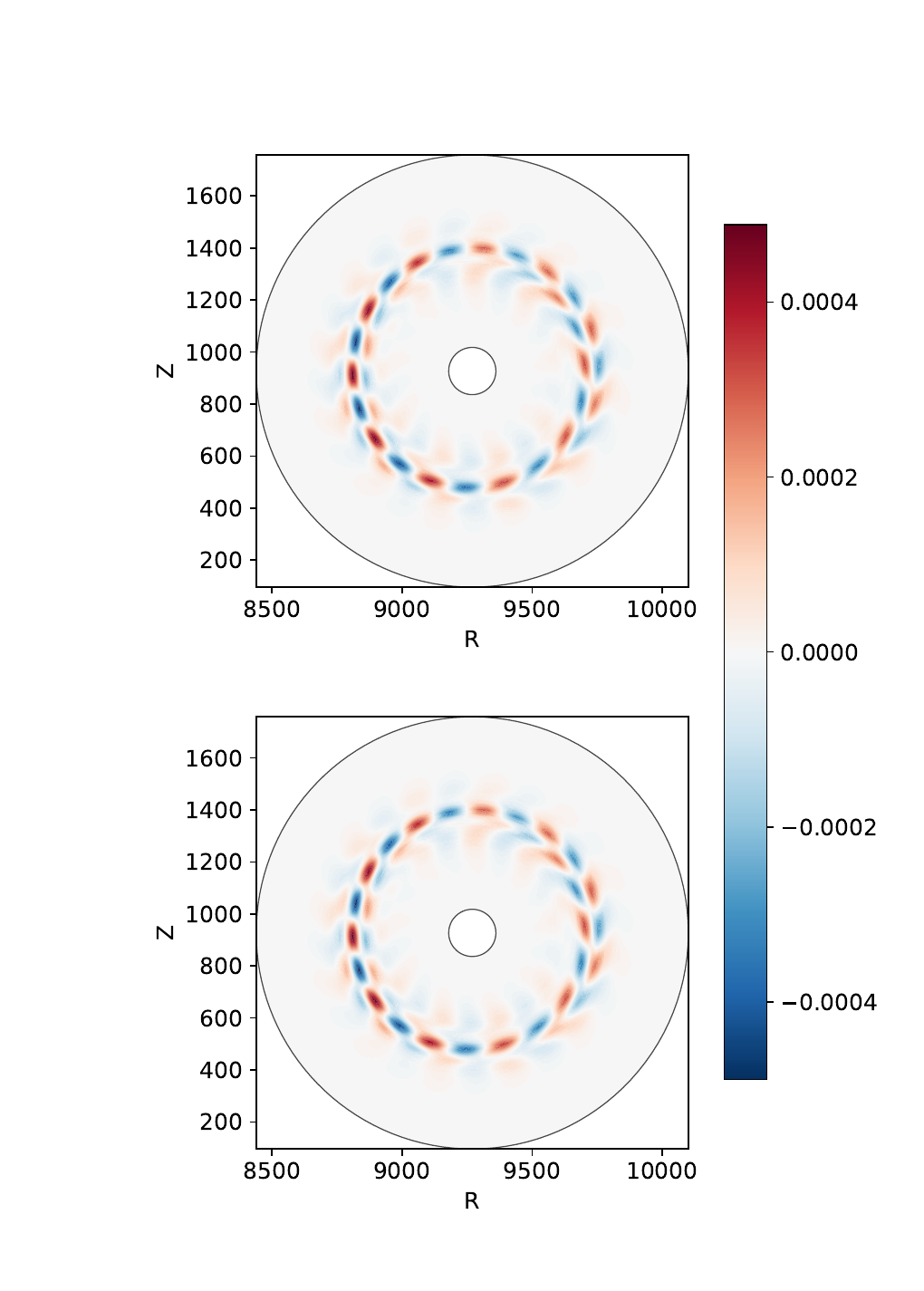}
    \\
\includegraphics[width=0.35\linewidth, trim=0 0 0cm 0cm, clip]{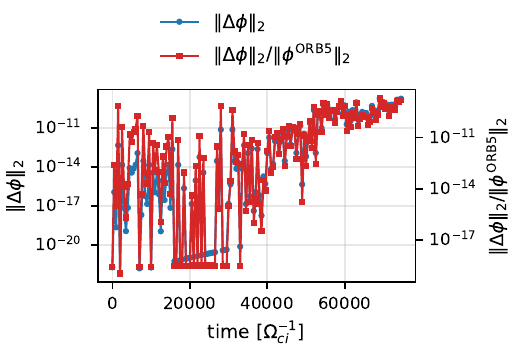}
    &
\includegraphics[width=0.35\linewidth, trim=0 0 0cm 0cm, clip]{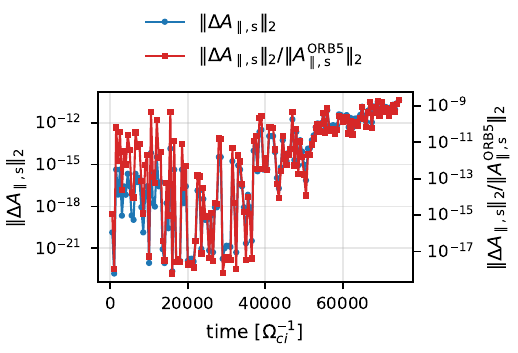}
&
\includegraphics[width=0.35\linewidth, trim=0 0 0cm 0cm, clip]{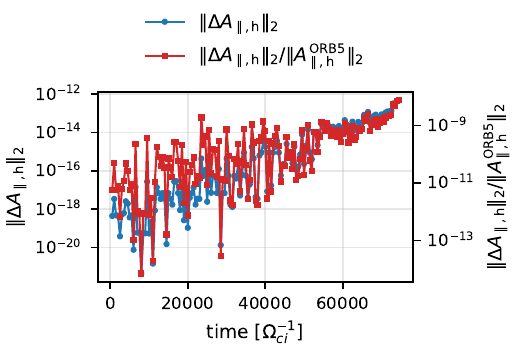}
    \end{tabular}
    \caption{ITPA final-time field comparison between ORB5X (top) and the ORB5 reference (middle) at $t=74500\ \Omega_{c_i}^{-1}$, as well as the evolution of error and relative error between them (bottom). The three panels show the electrostatic potential $\phi$, the symplectic part of the parallel vector potential $A_{||}^s$, and the Hamiltonian part of the parallel vector potential $A_{||}^h$.}
    \label{fig:potentials_final_time}
\end{figure}

\begin{figure}
\hspace{-2cm}
\centering
\begin{tabular}{cc}
\includegraphics[width=0.52\linewidth]{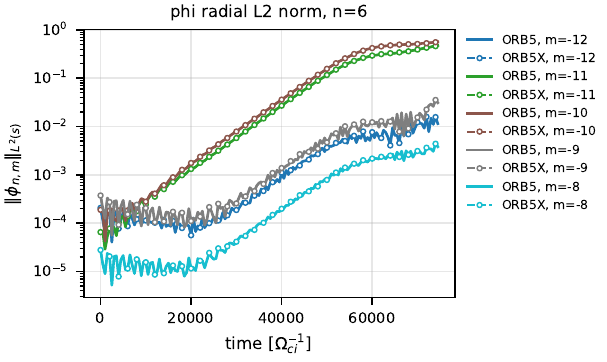}
    &
\includegraphics[width=0.52\linewidth]{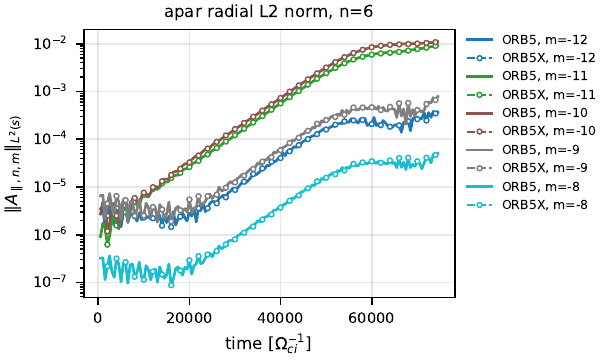}
\\
\includegraphics[width=0.52\linewidth]{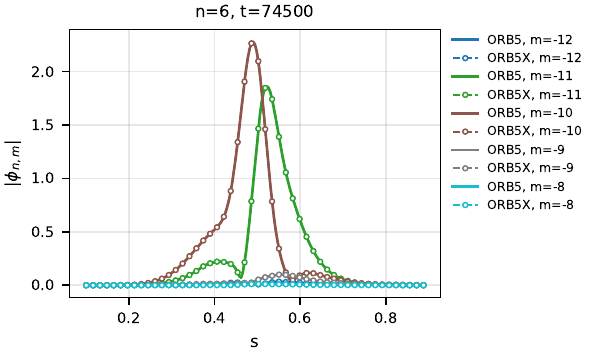}
    &
\includegraphics[width=0.52\linewidth]{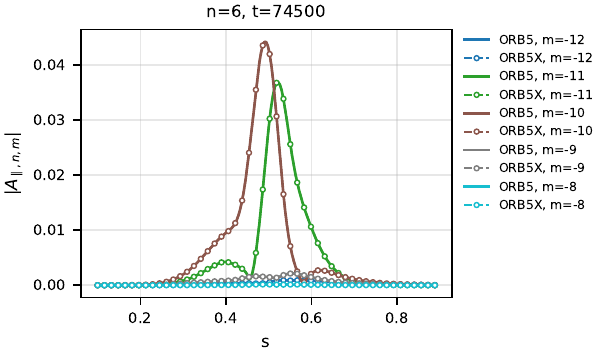}
        \end{tabular}
    \caption{Time evolution of ITPA Fourier modes obtained used ORB5 against ORB5X (top) as well as the radial mode structure at final time $t=74500\ \Omega_{c_i}^{-1}$ (bottom).}
    \label{fig:error_time}
\end{figure}

\begin{figure}
\hspace{-2cm}
    \centering
    \begin{tabular}{ccc}
\includegraphics[width=0.35\linewidth, trim=0 0 0cm 0cm, clip]{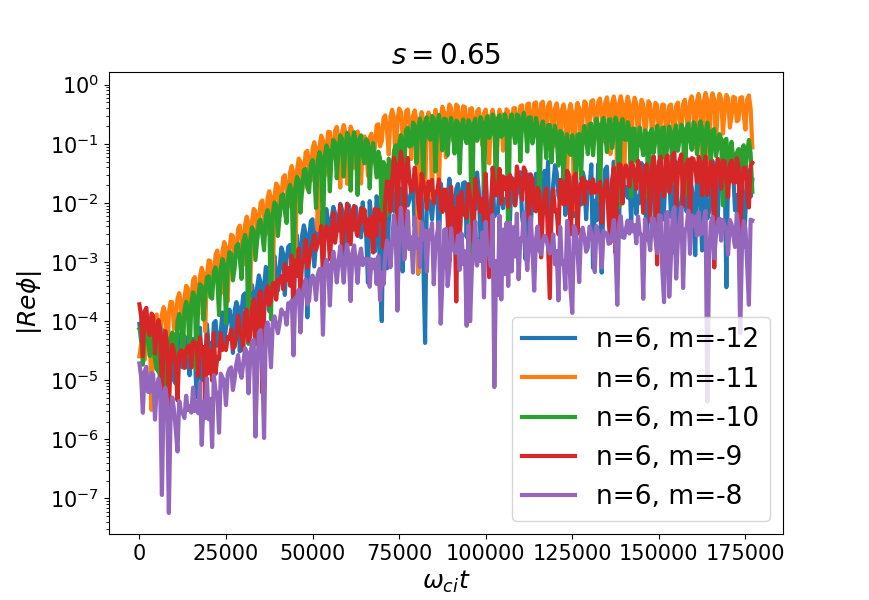}
    &
\includegraphics[width=0.35\linewidth, trim=0 0 0cm 0cm, clip]{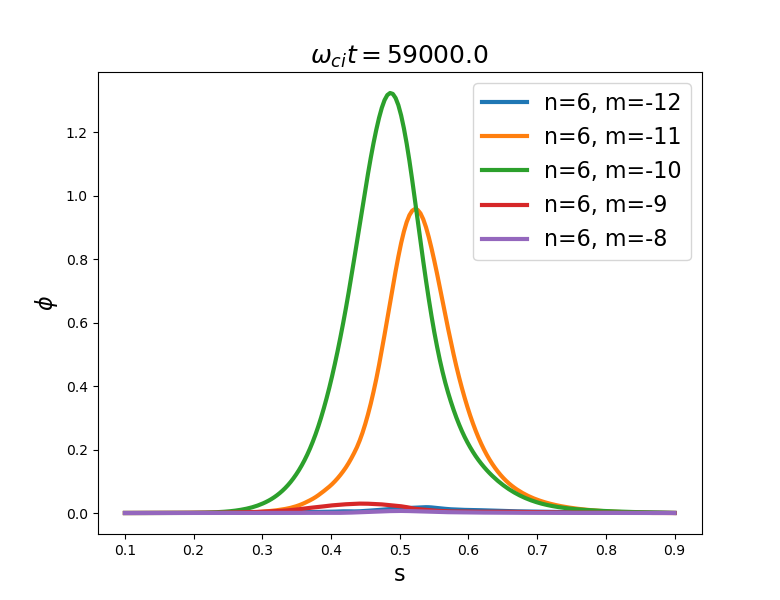}
    &
    \includegraphics[width=0.35\linewidth, trim=0 0 0cm 0cm, clip]{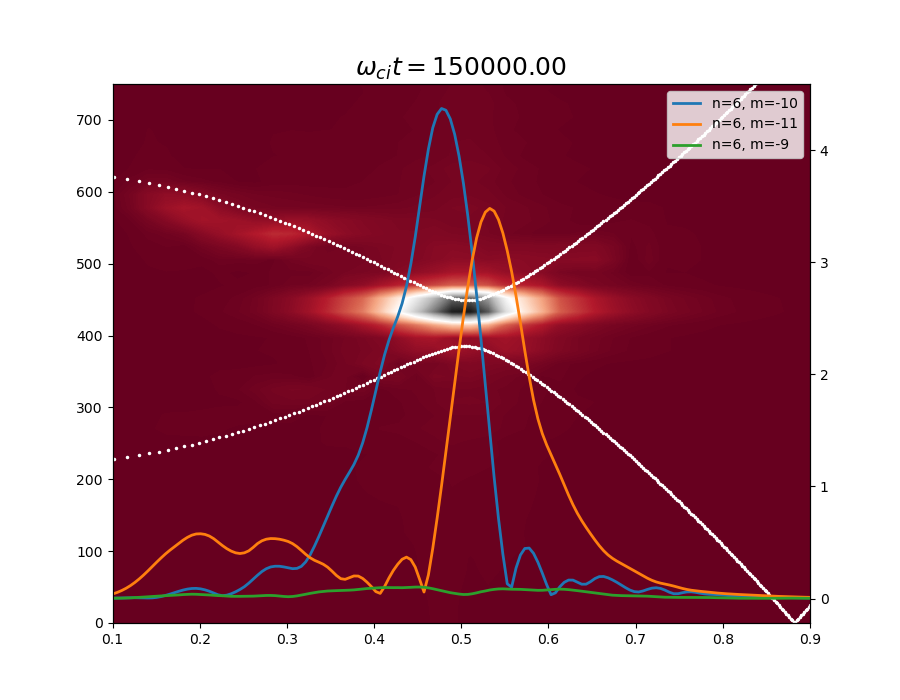}
        \end{tabular}
    \caption{(left) Saturation of modes due to the nonlinear interaction of the TAE wave and energetic particles, (middle) Formation of the TAE mode structure with the dominant poloidal harmonics $m=10$ and $m=11$, (right) Frequency of TAE located in the appropriate gap of the shear Alfv\'en continuum, obtained using ORB5X.}
    \label{fig:itpa-tae-analysis}
\end{figure}

The setup uses the ad-hoc circular ITPA equilibrium with $R_0=10$, $a=1$,
$B_0=3$, $\beta=4.48\times 10^{-4}$, and a coefficient-based safety-factor
profile.  The electromagnetic run contains kinetic deuterium ions, kinetic
electrons, and kinetic fast ions; the fast-ion density fraction is
$n_{0f}=0.0031$, with profile parameters $\kappa_n=3.333$, width $0.2$, and
peak position $s=0.5$.  The marker population is set by the input parameter
\texttt{nptot}: $10^7$ deuterium markers, $4\times10^7$ electron markers, and
$10^8$ fast-ion markers.  The fields are represented on a
$256\times256\times64$ $(s,\theta,\varphi)$ grid with cubic B splines
(\texttt{nidbas=3}), a sawtooth mode filter retaining the $n=6$,
$m=-9,\ldots,-12$ harmonics, the electromagnetic gyro-\(A_\parallel\) Ampere
solver, long-wavelength quasineutrality, and Ohm-law pullback.  
\\ \ \\
The comparison
in Fig.~\ref{fig:potentials_final_time}-\ref{fig:error_time} 
shows machine accuracy consistency between ORB5X and the original ORB5 in the evolution of $\phi$, $A_\parallel^s$, and
$A_\parallel^h$ for the same benchmark input. Furthermore, Fig.~\ref{fig:itpa-tae-analysis} shows the long term solution of ORB5X that correctly captures the nonlinear interaction of TAE wave and energetic particles with outcome frequency appearing the in the shear Alfv\'en continuum gap.

\FloatBarrier

\subsection{Chirping}

The Chirping case follows the energetic-particle-driven Alfvén-mode scenario
used to study nonadiabatic frequency chirping in ORB5
\cite{Wang2023Chirping}.  In this class of simulations, the nonlinear wave
frequency changes in time as the mode exchanges energy with resonant energetic
particles and the associated phase-space structures evolve.  The verification
case used here keeps that nonlinear electromagnetic setting, but compares the
ORB5X field evolution directly against the Fortran ORB5 reference for the same
input.

\begin{figure}
    \centering
    \begin{tabular}{@{}ccc@{}}
        \hspace{0.5cm} $\phi$ & $A_{||}^s$ & $A_{||}^h$
    \\
\includegraphics[width=0.315\linewidth, trim=0cm 0 0cm 2cm, clip]{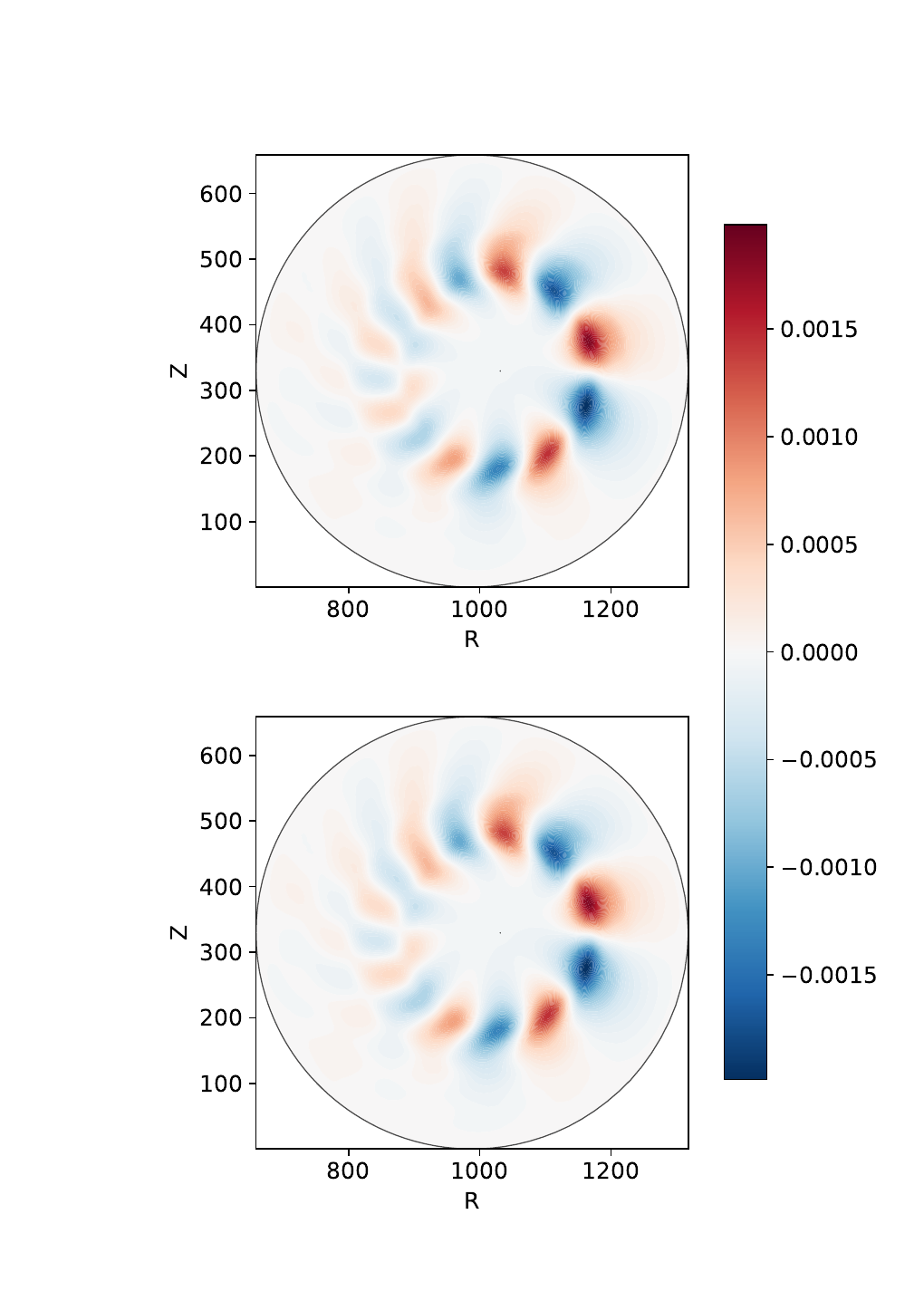}
    &
\includegraphics[width=0.315\linewidth, trim=0cm 0 0cm 2cm, clip]{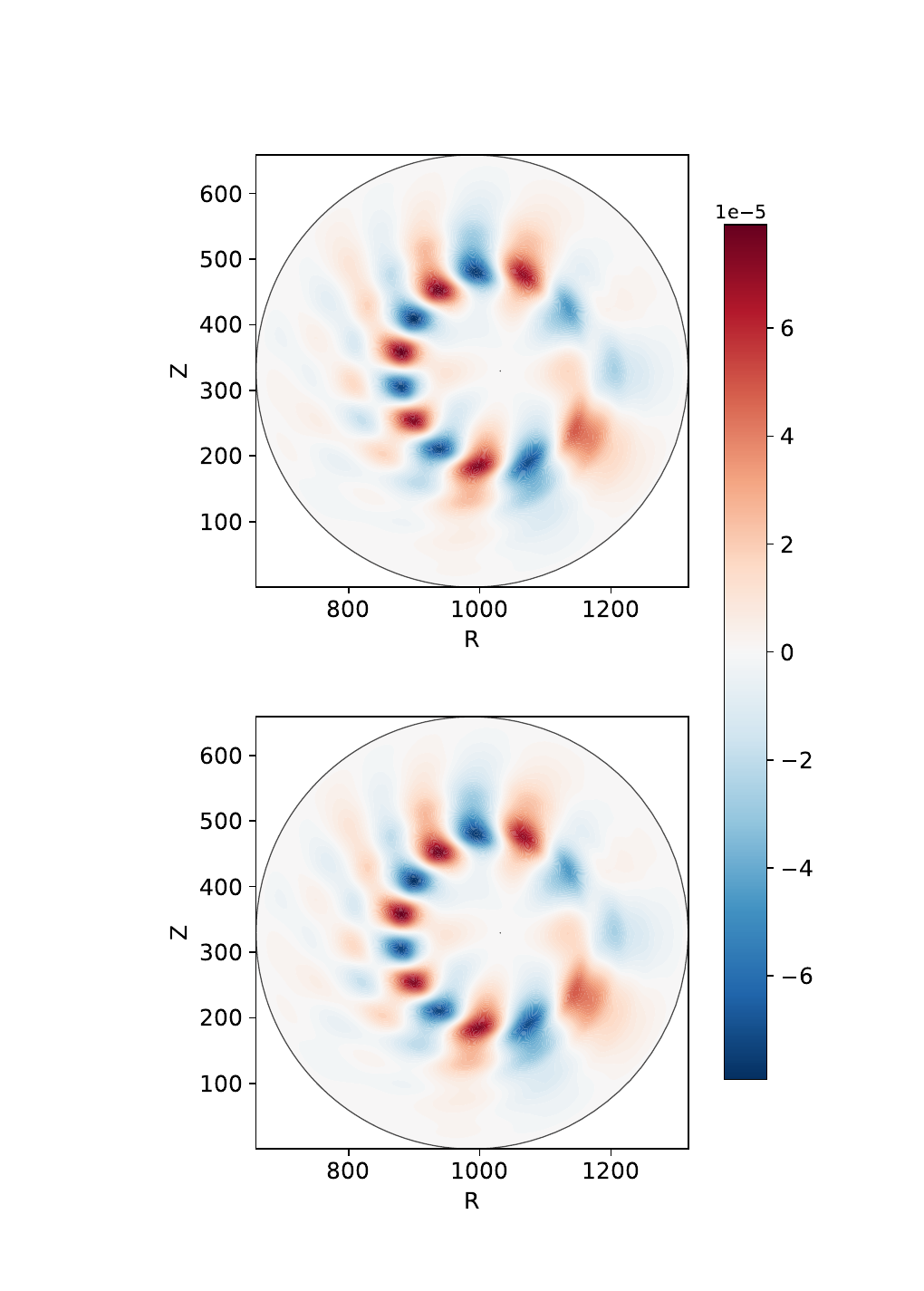}
    &
    \includegraphics[width=0.315\linewidth, trim=0cm 0 0cm 2cm, clip]{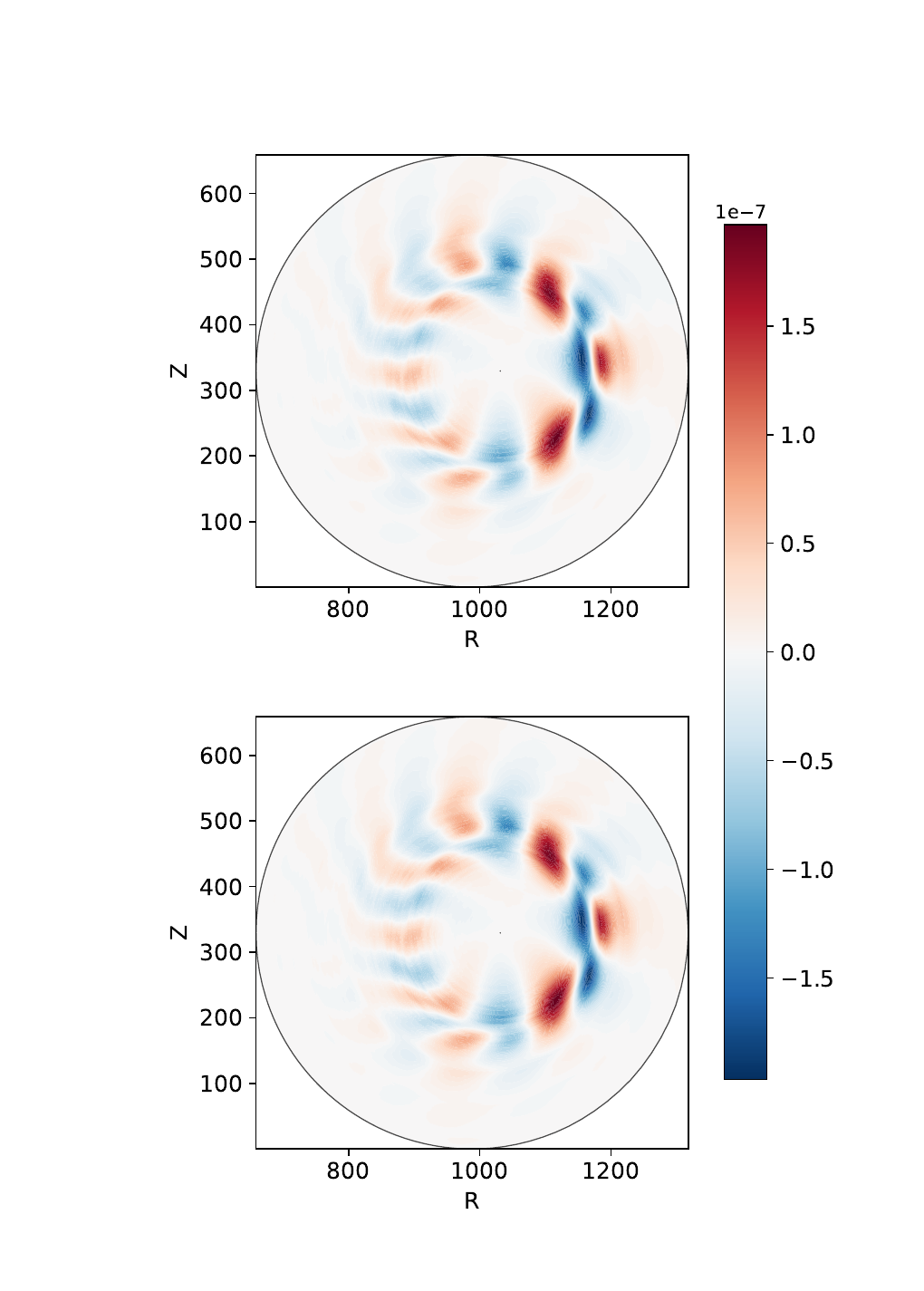}
    \\
    \includegraphics[width=0.315\linewidth, trim=0 0 0cm 0cm, clip]{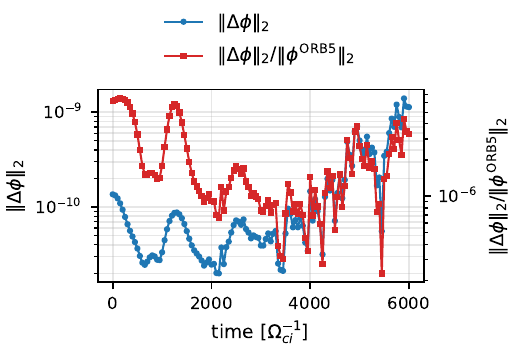}
    &
\includegraphics[width=0.315\linewidth, trim=0 0 0cm 0cm, clip]{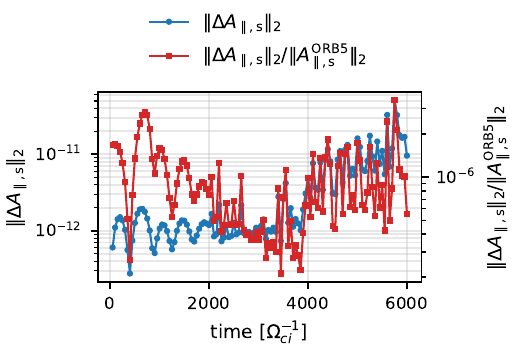}
    &
    \includegraphics[width=0.315\linewidth, trim=0 0 0cm 0cm, clip]{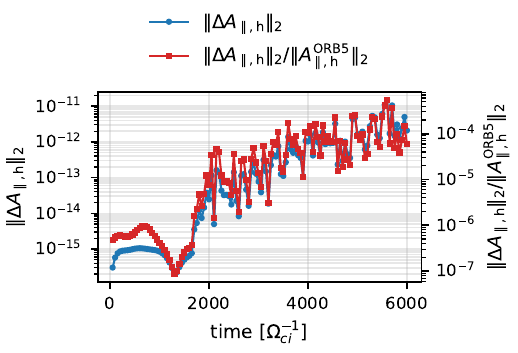}
    \end{tabular}
    \caption{Final-time field of Chirping tests cases obtained using ORB5X (top) and the reference ORB5 (middle) at final simulation time $t=6850\ \Omega_{c_i}^{-1}$, and the evolution of error and relative error (bottom). 
    }
    \label{fig:potentials_final_time_chirping}
\end{figure}

\begin{figure}
    \centering
    \includegraphics[width=0.975\linewidth, trim=0 0 0cm 0cm, clip]{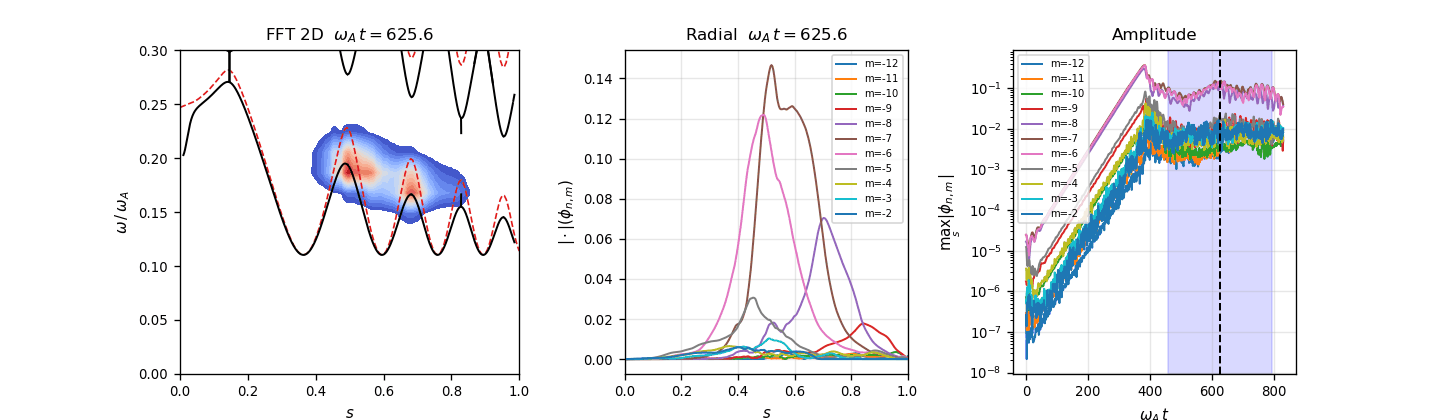}
    \caption{Chirping instability computed with ORB5X: frequency evolution plotted against the shear Alfv\'en continuum, radial mode structure and mode evolution in time, including nonlinear saturation of the instability. One sees that the physically correct behavior is reproduced with ORB5X, the same as seen with the Fortran version.}
    \label{fig:chirping_figure}
\end{figure}

The input uses an input MHD equilibrium
with
$\beta=4.4293\times10^{-3}$ and asymmetric-equilibrium terms enabled.  The
simulation has three kinetic species: thermal deuterium ions, kinetic
electrons, and energetic ions.  The deuterium and electron profiles are read
from the interpolated ITM profile tables with $\tau_i=\tau_e=1$, while the fast
ions use the same profile interface with $\tau_f=120$ and density fraction
$f_f=0.003$.  The corresponding input marker counts are
$6\times10^7$ deuterium markers, $1.2\times10^8$ electron markers, and
$6\times10^7$ energetic-ion markers.  The nonlinear electromagnetic fields are represented on a
$200\times128\times64$ $(s,\theta,\varphi)$ grid with cubic B splines, an
field-aligned \texttt{mn} Fourier filter over $n=0,\ldots,3$ with
$\Delta m=5$, and a flux-tube filter retaining multiples of $n=3$.  The initial
perturbation is a LIGKA-like automatic initialization centered on the
$(n,m)=(3,6)$ harmonic with width $0.1$ and amplitude $10^{-7}$.  The solver
uses the electromagnetic gyro-\(A_\parallel\) Ampere equation,
long-wavelength quasineutrality, and nonlinear Ohm-law pullback; phase-space
zonal-structure diagnostics are enabled in offline mode for the marker weights.
\\ \ \\
The comparison in Fig.~\ref{fig:potentials_final_time_chirping} shows 2D poloidal cross section of $\phi$, $A_\parallel^s$, and $A_\parallel^h$ distributions at
$t=6850\ \Omega_{c_i}^{-1}$ and the evolution of ORB5X error against ORB5 simulation. Again, we observe machine accuracy between ORB5X and the original ORB5 results. The physics of the chirping TAE instability, perfectly reproduced with ORB5X, is shown in Fig.~\ref{fig:chirping_figure}.

\FloatBarrier

\section{Performance}

\begin{figure}
    \centering
    \includegraphics[width=0.9\linewidth]{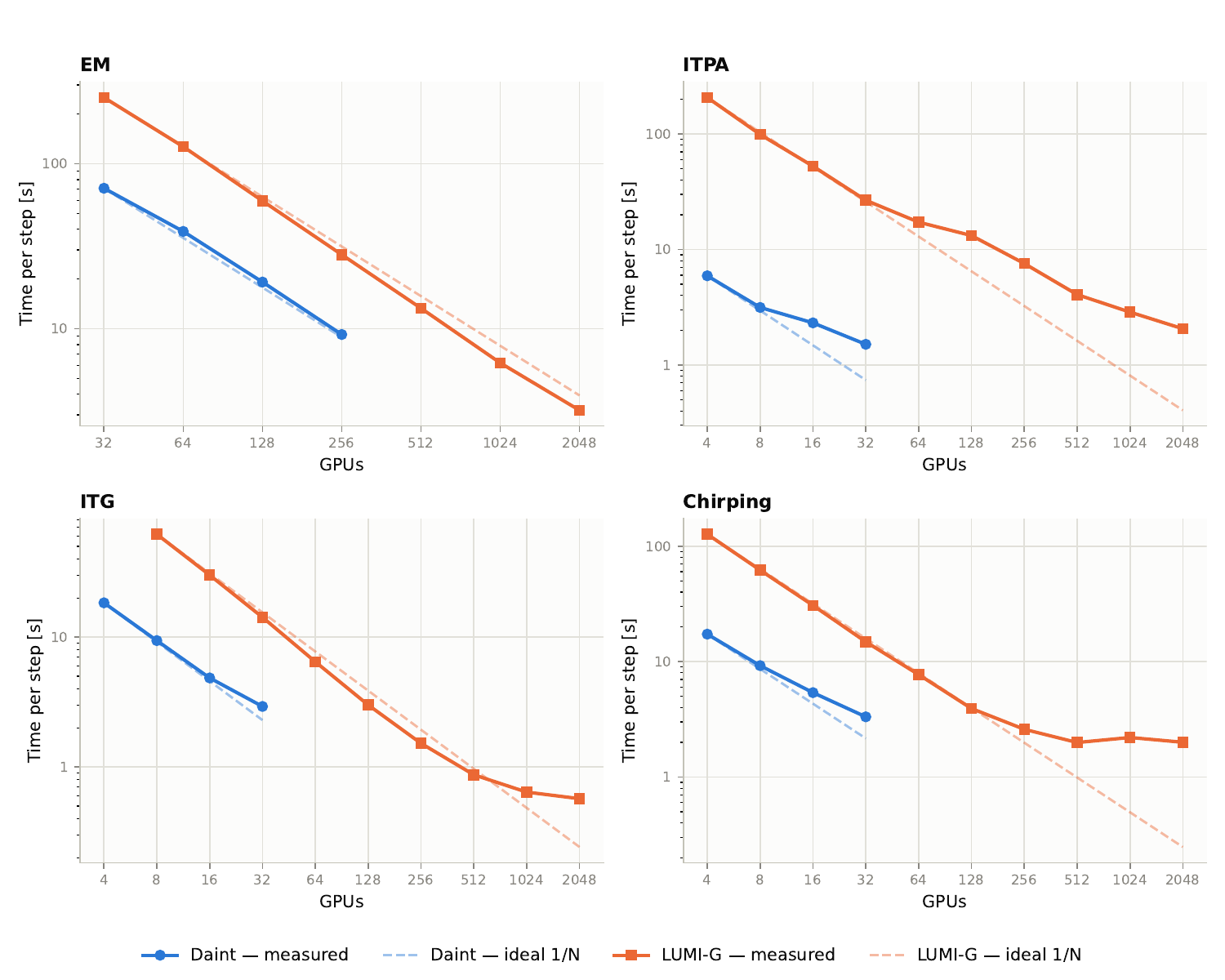}
    \caption{Strong scaling of ORB5X for full electromagnetic turbulence simulation (EM), ITPA, ITG, and Chirping test cases.}
    \label{fig:timing}
\end{figure}

We perform scaling studies of ORB5X using the validation test cases described in the previous sections, as well as an electromagnetic turbulence simulation (EM) with $4\times10^8$ deuterium, $10^9$ electron, and $2\times 10^8$ fast ion markers, with a mesh resolution of $256 \times 384 \times 192$ and a Fourier filter  that keeps modes within  $m \in  [-128, 128]$ and $n \in [0, 40]$ with $\Delta m=5$. The EM test  case represents the most computationally demanding and physically relevant test case considered in this work. As shown in Fig.~\ref{fig:timing}, ORB5X exhibits near-ideal scaling to large numbers of AMD MI250X GPUs on LUMI-G and NVIDIA GH200 GPUs on Daint at CSCS. Additional scaling studies demonstrating the performance portability of ORB5X across different HPC systems are presented in Appendix~\ref{sec:additional-portability-results}.

\FloatBarrier

\section{Conclusion}

\orbx{} is an initial C++17/Kokkos implementation of the ORB5 with MPI distributed memory parallelism, 
Kokkos shared-memory/device kernels, CMake-based dependency management, 
and a growing Fortran-parity test suite. 
The physical, mathematical and numerical model remains that of \orbfortran{}, 
i.e. global electromagnetic gyrokinetic PIC in toroidal geometry with spline finite elements, 
Fourier field solves, particle-field interpolation, collisions, noise control, and HDF5 diagnostics. 
The first ITG, ITPA, and Chirping comparisons show machine accuracy agreement
 for $\phi$, $A_{||}^h$ and $A_{||}^s$ with reasonable performance compared to the original ORB5. Furthermore, ORB5X is able to seamlessly be deployed on new HPC resources, reducing the operational costs associated with maintenance of the original Fortran code ORB5.
In the next works, we will focus on 
extending ORB5X to Stellarator physics by creating a flexible interface with VMEC++ \cite{schilling2025numerics}, EUTERPE \cite{sanchez2020nonlinear}, and general-geometry MHD solvers.

\section*{Acknowledgements}

MS acknowledges the financial support by the Paul Scherrer Institute and by QBIT Capital through
 the Founder Fellowship. 
  MS thanks Andreas Adelmann and Paolo Ricci for supporting this project.
  The numerical test cases were run using computational resources provided
    by the PSI's Merlin7 on ALPS,  Pitagora at CINECA, and CSCS Daint on ALPS. We acknowledge the EuroHPC Joint Undertaking for awarding this project access to the Discoverer+ supercomputer, hosted by Discoverer and LUMI-G hosted by LUMI through EuroHPC bechmkark call (Project ID: EHPC-BEN-2026B07-016).

\bibliographystyle{plain}
\bibliography{refs}

\appendix

\section{Detailed Physical Model}
\label{app:physical_model}

This appendix makes the ORB5X model description self-contained at the level needed to understand the translation and validation. The implementation follows a variational electromagnetic gyrokinetic model in the canonical-parallel-momentum, or $p_z$, formulation \cite{Sugama2000,Tronko2016,Brizard2007}. The ordering distinguishes the magnetic-geometry parameter $\epsilon_B=\rho_{\mathrm{th}}/L_B$ from the fluctuation parameter $\epsilon_\delta$, with the global-code ordering $\epsilon_B=\epsilon_\delta^2$. The perpendicular fluctuation scale is measured through $\epsilon_\perp\sim k_\perp\rho_{\mathrm{th}}$, allowing both gyrokinetic and drift-kinetic limits in the same framework.

\subsection{Geometry, Coordinates, and Normalization}

ORB5X uses the same axisymmetric toroidal equilibrium description as ORB5. In the region of nested
flux surfaces the equilibrium magnetic field is written
\begin{equation}
  \bm{B} = F(\psi)\nabla\varphi + \nabla\psi\times\nabla\varphi ,
\end{equation}
where $\psi$ is the poloidal magnetic flux, $\varphi$ is the toroidal angle, and $F(\psi)$ is the
poloidal-current flux function. Equilibria may be read from ideal-MHD Grad--Shafranov solutions, for
example CHEASE equilibria \cite{Lutjens1996}, or supplied by analytic circular concentric surfaces.
The magnetic-surface label used throughout the code is
\begin{equation}
  s=\sqrt{\psi/\psi_{\mathrm{edge}}},
\end{equation}
and the straight-field-line poloidal angle is defined by
\begin{equation}
  \chi =
  \frac{1}{q(s)}\int_0^\theta
  \frac{\bm{B}\cdot\nabla\varphi}{\bm{B}\cdot\nabla\theta'}\,d\theta' ,
\end{equation}
where $q(s)$ is the safety-factor profile and $\theta$ is the geometric poloidal angle. The marker
push is therefore formulated in magnetic coordinates $(s,\chi,\varphi)$, except near the magnetic
axis where ORB5X follows ORB5 and switches to $(s\cos\chi,s\sin\chi,\varphi)$ to avoid coordinate
singularities.

The internal normalization is fixed by four reference quantities: a reference ion mass $m_i$, a
reference ion charge $q_i=eZ_i$, the magnetic-field magnitude at the magnetic axis $B_0$, and the
electron temperature $T_e(s_0)$ on a reference surface. Time is normalized to
$\Omega_{ci}^{-1}$ with $\Omega_{ci}=q_iB_0/(m_i c)$, velocities to the sound speed
$c_s=\sqrt{eT_e(s_0)/m_i}$, lengths to $\rho_s=c_s/\Omega_{ci}$, and densities to the volume-averaged
density. These conventions are inherited from ORB5 so that input parameters, diagnostic quantities,
and validation comparisons can be interpreted without changing physical units.

\subsection{Variational Model and Field Equations}

The field-particle action used by the model can be written schematically as
\begin{align}
  \mathcal{A}=\int_{t_0}^{t_1} dt\,\mathcal{L}
  =&
  \sum_s\int dt\,d\Omega
  \left[
    \frac{q_s}{c}\bm{A}^*\cdot\dot{\bm{X}}
    +\frac{m_s c}{q_s}\mu\dot{\theta}
    -H_0
  \right]f_s \nonumber\\
  &-\epsilon_\delta\sum_{s\neq e}\int dt\,d\Omega\,H_1 f_s
  -\epsilon_\delta\int dt\,d\Omega\,H_1^{\mathrm{dk}} f_e \nonumber\\
  &-\epsilon_\delta^2\sum_{s\neq e}\int dt\,d\Omega\,H_2 f_{\mathrm{eq},s}
  -\alpha\epsilon_\delta^2\int dt\,d\Omega\,H_2^{\mathrm{dk}} f_{\mathrm{eq},e}
  -\alpha\epsilon_\delta^2\int dt\,dV\,\frac{|\nabla_\perp A_{1\parallel}|^2}{8\pi}.
\end{align}
Here $d\Omega=dV\,dW$, $dW=B_\parallel^*\,d\mu\,dp_z$, $\bm{B}^*=\nabla\times\bm{A}^*$, and $\bm{A}^*=\bm{A}+(c/q_s)p_z\bm{b}$. The quasineutrality approximation neglects the perturbed electric-field energy, magnetic compressibility is neglected so that the perpendicular perturbed magnetic field is represented by $\bm{B}_{1\perp}=\bm{b}\times\nabla A_{1\parallel}$, and the second-order Hamiltonian terms are evaluated on equilibrium distributions. These choices make the field equations linear in the unknown fields while preserving the nonlinear particle response through the characteristics and weights.

The gyrocenter phase-space variables are the position $\bm{X}$, magnetic moment $\mu$, gyroangle $\theta$, and
\begin{equation}
  p_z = m_s v_\parallel + \alpha\,\epsilon_\delta \frac{q_s}{c} A_{1\parallel},
  \qquad
  \alpha =
  \begin{cases}
  0, & \text{electrostatic model},\\
  1, & \text{electromagnetic model}.
  \end{cases}
\label{eq:pz_definition}
\end{equation}
Here $q_s$ and $m_s$ are the charge and mass of species $s$, $A_{1\parallel}$ is the perturbed parallel vector potential, and $\epsilon_\delta$ is the perturbation-ordering parameter. The lowest-order Hamiltonian is
\begin{equation}
  H_0 = \frac{p_z^2}{2m_s}+\mu B .
\end{equation}
The first-order particle-field coupling for gyrokinetic ions is
\begin{equation}
  H_1 = q_s\left\langle \phi_1-\alpha A_{1\parallel}\frac{p_z}{m_s c}\right\rangle,
\label{eq:first_order_hamiltonian}
\end{equation}
where $\langle\cdot\rangle$ denotes gyroaveraging over the Larmor orbit. For drift-kinetic electrons the corresponding coupling is
\begin{equation}
  H_1^{\mathrm{dk}} =
  -e\left(\phi_1-\alpha A_{1\parallel}\frac{p_z}{m_e c}\right).
\end{equation}
Second-order Hamiltonian terms provide the polarization and magnetization contributions to the field equations. ORB5X retains the ORB5 choices for full-FLR electrostatic polarization, long-wavelength polarization, and Pad\'e-like approximations where those paths are translated.

The nonlinear gyrokinetic Vlasov equation is obtained by varying the action with respect to the distribution function. In practical PIC form, ORB5X advances gyrocenter characteristics generated by the selected Hamiltonian model and evolves marker weights associated with $f_0$ and $\delta f$. The equilibrium distribution entering the second-order field terms satisfies $\{f_{\mathrm{eq},s},H_0\}=0$, while the simulation control variate $f_{0s}$ is the chosen numerical background distribution used by the $\delta f$ method.
\begin{equation}
  0=\frac{df_s}{dt}
  =
  \frac{\partial f_s}{\partial t}
  +\dot{\bm{X}}\cdot\nabla f_s
  +\dot{p}_z\frac{\partial f_s}{\partial p_z}.
\label{eq:vlasov_characteristic_form}
\end{equation}
For a Hamiltonian $H=H_0+\epsilon_\delta H_1$, the characteristics are
\begin{align}
  \dot{\bm{X}}
  &=
  \frac{c\bm{b}}{q_s B_\parallel^*}\times\nabla H
  +\frac{\partial H}{\partial p_z}\frac{\bm{B}^*}{B_\parallel^*},\\
  \dot{p}_z
  &=
  -\frac{\bm{B}^*}{B_\parallel^*}\cdot\nabla H .
\label{eq:hamiltonian_characteristics}
\end{align}
Substituting the ORB5 electromagnetic Hamiltonian gives
\begin{align}
  \dot{\bm{X}}
  &=
  \frac{c\bm{b}}{q_s B_\parallel^*}\times
  \nabla\left[
    \mu B+\epsilon_\delta q_s
    \left(\langle\phi_1\rangle-\alpha\frac{p_z}{m_s c}\langle A_{1\parallel}\rangle\right)
  \right]
  +\frac{\bm{B}^*}{B_\parallel^*}
  \left[
    \frac{p_z}{m_s}
    -\epsilon_\delta\alpha\frac{q_s}{m_s c}\langle A_{1\parallel}\rangle
  \right],\\
  \dot{p}_z
  &=
  -\frac{\bm{B}^*}{B_\parallel^*}\cdot
  \nabla\left[
    \mu B+\epsilon_\delta q_s
    \left(\langle\phi_1\rangle-\alpha\frac{p_z}{m_s c}\langle A_{1\parallel}\rangle\right)
  \right].
\label{eq:em_characteristics}
\end{align}
These equations contain the parallel motion, magnetic drifts, curvature drift, $E\times B$ drift,
and electromagnetic $A_\parallel$ corrections. In the linear limit the terms proportional to
$\epsilon_\delta$ are omitted in the characteristics; in the neoclassical limit the fluctuation
fields are omitted and the magnetic-drift terms are further neglected relative to the parallel
motion. These switches are algorithmic choices in ORB5X rather than changes to the field
discretization.

The distribution function is represented by numerical markers in $\delta f$ form,
\begin{equation}
  f_s = f_{0s}+\delta f_s .
\label{eq:delta_f_decomposition}
\end{equation}
The control variate $f_{0s}$ may be a local Maxwellian, canonical Maxwellian, or corrected-canonical Maxwellian. A local Maxwellian $f_L(\psi,\epsilon,\mu)$ matches prescribed density and temperature profiles directly, but because $\psi$ is not an exact invariant of the unperturbed toroidal motion it can drive spurious zonal-flow response if the unperturbed contribution to $df_0/dt$ is retained. A canonical Maxwellian $f_C(\psi_0,\epsilon,\mu)$ is an equilibrium distribution because it depends on constants of motion, but its effective profiles can differ from the requested profiles. ORB5 therefore also uses a corrected canonical variable
\begin{equation}
  \widehat{\psi}
  =
  \psi_0
  -\operatorname{sign}(v_\parallel)\frac{m_s c}{q_s}R_0
  \sqrt{2(\epsilon-\mu B_0)}\,\mathcal{H}(\epsilon-\mu B_0),
\end{equation}
where $\mathcal{H}$ is the Heaviside function. The correction vanishes for trapped particles and
has opposite sign for co- and counter-passing particles, reducing the profile mismatch while keeping
$f_{CC}(\widehat{\psi},\epsilon,\mu)$ an equilibrium distribution. ORB5X preserves the ORB5 marker decomposition: time-dependent coordinates and weights are stored separately from quantities constant along the orbit, including $\mu$, initial phase-space volume, initial distribution values, labels, and auxiliary reduced-weight quantities. This layout is a physics-level invariant because particle push, diagnostics, collisions, pullback, and particle migration all assume the same marker attributes.

The weak quasineutrality equation balances gyroaveraged ion charge, drift-kinetic electron charge, and polarization charge. With test function $\widehat{\phi}_1$ the structure is
\begin{equation}
  \sum_{s\neq e} Q_s^{\mathrm{gyr}}(\widehat{\phi}_1)
  +Q_e^{\mathrm{dk}}(\widehat{\phi}_1)
  =
  \sum_{s\neq e} Q_s^{\mathrm{pol}}(\widehat{\phi}_1).
\end{equation}
The full-FLR form used by the parent ORB5 model can be written as
\begin{align}
  Q_s^{\mathrm{gyr}}
  &=
  \int d\Omega\,f_s q_s\langle\widehat{\phi}_1\rangle,\\
  Q_e^{\mathrm{dk}}
  &=
  -\int d\Omega\,f_e e\,\widehat{\phi}_1(\bm{X}),\\
  Q_s^{\mathrm{pol}}
  &=
  \epsilon_\delta
  \int d\Omega\,f_{\mathrm{eq},s}
  \frac{q_s^2}{B}
  \frac{\partial}{\partial\mu}
  \left[
    \langle\phi_1\widehat{\phi}_1\rangle
    -\langle\phi_1\rangle\langle\widehat{\phi}_1\rangle
  \right].
\end{align}
In the long-wavelength limit, the ion polarization term contains
\begin{equation}
  \int d\Omega\, f_{\mathrm{eq},s}
  \frac{m_s c^2}{B^2}
  \nabla_\perp\phi_1\cdot\nabla_\perp\widehat{\phi}_1 .
\end{equation}
The weak Amp\`ere equation contains perpendicular magnetic energy, ion and electron parallel-current projections, skin terms, and ion magnetic-FLR corrections:
\begin{align}
0
&=
-\epsilon_\delta\int \frac{dV}{4\pi}
\nabla_\perp A_{1\parallel}\cdot\nabla_\perp\widehat{A}_{1\parallel}
\sum_{s\neq e}\int d\Omega\,f_s
\frac{q_s p_z}{m_s c}\langle\widehat{A}_{1\parallel}\rangle
-\int d\Omega\,f_e\frac{e p_z}{m_e c}\widehat{A}_{1\parallel}
\nonumber\\
&\quad
-\epsilon_\delta\int d\Omega\,f_{\mathrm{eq},e}
\frac{e^2}{m_e c^2}A_{1\parallel}\widehat{A}_{1\parallel}
\nonumber\\
&\quad
-\epsilon_\delta\sum_{s\neq e}\int d\Omega\,f_{\mathrm{eq},s}
\left[
  \frac{q_s^2}{m_s c^2}A_{1\parallel}\widehat{A}_{1\parallel}
  +\frac{\mu}{2B}
  \left(
    A_{1\parallel}\nabla_\perp^2\widehat{A}_{1\parallel}
    +\widehat{A}_{1\parallel}\nabla_\perp^2A_{1\parallel}
  \right)
\right].
\end{align}
In adiabatic-electron simulations, the kinetic electron response is replaced by a fluid response proportional to $\phi_1-\overline{\phi}_1$, where $\overline{\phi}_1$ is the flux-surface average. Hybrid-electron variants combine kinetic and fluid electron pieces.

\subsection{Model Variants}

The long-wavelength approximation replaces the all-order electrostatic FLR part of $H_2$ by its
second-order expansion,
\begin{equation}
  H_2^{\mathrm{LWA}}
  =
  -\frac{m_s c^2}{2B^2}|\nabla_\perp\phi_1|^2
  +\alpha\frac{q_s^2}{2m_s c^2}
  \left[
    A_{1\parallel}^2
    +m_s\left(\frac{c}{q_s}\right)^2\frac{\mu}{B}
    A_{1\parallel}\nabla_\perp^2 A_{1\parallel}
  \right],
\end{equation}
which produces the long-wavelength polarization term already shown above and leaves the
first-order characteristics unchanged. The Pad\'e model rewrites the ion gyroaverage through the
operator $(1-\nabla_\perp\cdot\rho_i^2\nabla_\perp)^{-1}$ so that the inverse full matrix does not
have to be formed explicitly. In weak form, the quasineutrality equation is multiplied by
$(1-\nabla_\perp\cdot\rho_i^2\nabla_\perp)$, preserving a sparse finite-element structure while
approximating finite-Larmor-radius shielding.

For adiabatic electrons, the electron density perturbation is represented as a flux-surface-corrected Boltzmann response, so the zonal component is removed through the flux-surface average. The electron part of quasineutrality becomes
\begin{equation}
  Q_e^{\mathrm{ad}}
  =
  \epsilon_\delta\int dV\,
  \frac{e n_{e0}}{T_e}(\phi_1-\overline{\phi}_1)\widehat{\phi}_1
  -\int dV\,n_{e0}\widehat{\phi}_1 .
\end{equation}
For hybrid electrons, a fraction of passing electrons is treated adiabatically while the trapped
population remains drift kinetic. The corresponding electron contribution is
\begin{equation}
  Q_e^{\mathrm{hyb}}
  =
  \alpha_P\epsilon_\delta\int dV\,
  \frac{e n_{e0}}{T_e}(\phi_1-\overline{\phi}_1)\widehat{\phi}_1
  -\int dV\,n_{e0}\widehat{\phi}_1
  -e\int_{\mathrm{trapped}}d\Omega\,f_e\widehat{\phi}_1
  +e\int_{\mathrm{passing}}d\Omega\,f_e\widehat{\phi}_1^{00},
\end{equation}
where $\widehat{\phi}_1^{00}$ is the zonal $(n,m)=(0,0)$ component of the test function. In electromagnetic simulations the kinetic and field contributions to $A_\parallel$ are split into symplectic and Hamiltonian parts, and pullback operations transfer information between the marker weights and the field variables to mitigate cancellation error.

Electromagnetic simulations use a split representation of the parallel vector potential into symplectic and Hamiltonian components, with pullback operations used to reduce the cancellation problem \cite{Mishchenko2017,Hatzky2007,Hatzky2019,Mishchenko2018a}. This is one of the most sensitive model-preservation points in ORB5X: the time-step workflow must transfer information between particle weights, the split $A_\parallel$ components, and field solves without changing the control-variate algebra.

Collisions follow the same model family as the Fortran implementation \cite{Vernay2010}. The background distribution is converted to a local Maxwellian representation when needed. Self-collisions use a linearized Landau operator split into a test-particle drag/diffusion term and a background-reaction term. The test-particle term is applied through stochastic velocity-space kicks, while the background reaction restores conservation of density, parallel momentum, and energy in each spatial bin \cite{Lin1995a,Brunner1999}. Electron-ion collisions use a Lorentz pitch-angle-scattering operator in the large-mass-ratio limit.
For self-collisions the test-particle part has the drag-diffusion form
\begin{equation}
  C[f_L,\delta f]
  =
  \frac{\partial}{\partial\bm{v}}\cdot\left(\bm{\Gamma}\delta f\right)
  -\frac{\partial^2}{\partial\bm{v}\partial\bm{v}}:\left(\bm{D}\delta f\right),
\end{equation}
with
\begin{equation}
  \bm{\Gamma}=-\bar{\nu}H(x)\bm{v},
  \qquad
  \bm{D}=\frac{\bar{\nu}v_{\mathrm{th}}^2}{4}
  \left[
    K(x)\left(\bm{I}-\frac{\bm{v}\bm{v}}{v^2}\right)
    +2H(x)\frac{\bm{v}\bm{v}}{v^2}
  \right],
\end{equation}
where $x=v/(\sqrt{2}v_{\mathrm{th}})$ and $\bar{\nu}=8\pi n q^4\ln\Lambda/(m^2v_{\mathrm{th}}^3)$.
The background-reaction approximation is written
\begin{equation}
  C[\delta f,f_L]\simeq f_L\beta(\delta f),\qquad
  \beta(\delta f)
  =
  \frac{1}{n}
  \left[
    6\sqrt{\pi}H(x)\frac{\delta P_\parallel v_\parallel}{v_{\mathrm{th}}^2}
    +\sqrt{\pi}G(x)\frac{\delta E}{v_{\mathrm{th}}^2}
  \right],
\end{equation}
which is the term used to restore the moments lost or gained by the stochastic test-particle kicks.
The electron-ion operator retained in ORB5 is the Lorentz pitch-angle-scattering limit,
\begin{equation}
  C_{ei}[f_i,\delta f_e]
  =
  -\nu_{ei}(v)
  \frac{\partial}{\partial\xi}
  \left[
    (1-\xi^2)\frac{\partial\delta f_e}{\partial\xi}
  \right],
  \qquad
  \nu_{ei}(v)=\frac{\bar{\nu}_{ei}}{4}\left(\frac{v_{\mathrm{th},e}}{v}\right)^3 .
\end{equation}

The optional strong-flow model permits toroidal rotation with background $E\times B$ speed comparable to the ion thermal speed \cite{Hahm1996,Collier2016}. In that ordering,
\begin{equation}
  H_0^{\mathrm{flow}}
  =
  q_s\Phi+\mu B+\frac{p_z^2+m_s^2|\bm{u}_E|^2}{2m_s},
  \qquad
  \bm{u}_E=\frac{c}{B}\bm{b}\times\nabla\Phi .
\end{equation}
The corresponding canonical momentum includes the toroidal component of the background flow. ORB5X contains the translated strong-flow path and diagnostics, but the paper should claim production-level support only after a rotating-plasma validation case has been included.

The conservation laws used for diagnostics follow from the same variational structure. The electromagnetic energy separates into kinetic and field contributions. In the absence of external sources and numerical error, the power balance has the form
\begin{equation}
  \frac{d\mathcal{E}_{\mathrm{kin}}}{dt}
  =
  -\frac{d\mathcal{E}_{\mathrm{F}}}{dt}.
\end{equation}
The kinetic part is evaluated from particle quantities, while the field part is evaluated from grid quantities and the weak field equations. This separation is essential in a PIC code because particles and fields are represented on different numerical objects.
For the general electromagnetic model, the invariant energy before rewriting with the weak field
equations is
\begin{align}
  \mathcal{E}^{\mathrm{EM}}
  &=
  \sum_s\int d\Omega\,H_0 f_s
  +\epsilon_\delta\sum_{s\neq e}\int d\Omega\,H_1 f_s
  +\epsilon_\delta\int d\Omega\,H_1^{\mathrm{dk}} f_e
  \nonumber\\
  &\quad
  +\epsilon_\delta^2\sum_{s\neq e}\int d\Omega\,H_2 f_{\mathrm{eq},s}
  +\alpha\epsilon_\delta^2\int d\Omega\,H_2^{\mathrm{dk}}f_{\mathrm{eq},e}
  +\alpha\epsilon_\delta^2\int dV\,\frac{|\nabla_\perp A_{1\parallel}|^2}{8\pi}.
\end{align}
The kinetic diagnostic is normally taken from
\begin{equation}
  \mathcal{E}_{\mathrm{kin}}
  =
  \sum_s\int d\Omega
  \left(\frac{p_z^2}{2m_s}+\mu B\right)f_s ,
\end{equation}
while an equivalent field-energy expression can be obtained by choosing
$\widehat{\phi}_1=\phi_1$ and $\widehat{A}_{1\parallel}=A_{1\parallel}$ in the weak field
equations. In density-current form this yields
\begin{equation}
  \mathcal{E}_F
  =
  \frac{\epsilon_\delta}{2}\sum_{s\neq e}q_s\int d\Omega\,
  \left(
    \langle\phi_1\rangle-\alpha\frac{p_z}{m_s c}\langle A_{1\parallel}\rangle
  \right)f_s
  -\frac{\epsilon_\delta}{2}e\int d\Omega\,
  \left(
    \phi_1-\alpha\frac{p_z}{m_e c}A_{1\parallel}
  \right)f_e .
\end{equation}
This equality between particle-projected and field-projected energies is the basis of the ORB5 and
ORB5X power-balance diagnostics.

\begin{table}[H]
\centering
\scriptsize
\renewcommand{\arraystretch}{0.88}
\setlength{\tabcolsep}{2pt}
\begin{tabular}{p{0.1\linewidth}p{0.3\linewidth}p{0.27\linewidth}p{0.23\linewidth}}
\toprule
Symbol & Mathematical meaning & ORB5X representation & Numerical role \\
\midrule
$\bm{X}$ & Gyrocenter position. & Particle coordinate Views in the particle module. & Particle push, deposition, interpolation, migration. \\
$s$ & Radial flux coordinate, $s=\sqrt{\psi/\psi_{\mathrm{edge}}}$. & Marker coordinate and radial grid index. & Radial profiles, splines, filters, diagnostics. \\
$\chi$ & Straight-field-line poloidal angle. & Marker coordinate and poloidal grid index. & B-spline basis, Fourier modes, field-aligned filter. \\
$\varphi$ & Toroidal angle. & Marker coordinate and toroidal subdomain coordinate. & Domain decomposition, particle migration, toroidal FFT. \\
$\mu$ & Magnetic moment. & Constant marker attribute. & Hamiltonian, gyroaveraging, diagnostics. \\
$p_z$ & Canonical parallel momentum. & Particle momentum coordinate. & Gyrocenter characteristics and electromagnetic model. \\
$v_\parallel$ & Kinetic parallel velocity. & Derived from $p_z$ and $A_\parallel$ when needed. & Diagnostics, collisions, source terms. \\
$f_s$ & Full species distribution function. & Represented by markers and weights. & Statistical PIC representation of species. \\
$f_{0s}$ & Control-variate/background distribution. & Initial/background marker weight channel, profile modules. & $\delta f$ weights, direct-$\delta f$, collisions. \\
$\delta f_s$ & Perturbed distribution. & Marker perturbation weights. & Charge/current deposition and diagnostics. \\
$p_i$ & Marker weight for $f_0$. & Particle weight channel. & Standard and direct $\delta f$ algorithms. \\
$\delta w_i$ & Marker perturbation weight. & Particle weight channel. & Deposited density/current and field solve RHS. \\
$\phi$ & Electrostatic potential. & \texttt{phi.bspl}. & Quasineutrality solve and interpolation to particles. \\
$A_\parallel$ & Parallel vector potential. & \texttt{apar.bspl}. & Electromagnetic Amp\`ere solve and particle push. \\
$A_\parallel^s$ & Symplectic component of $A_\parallel$. & \texttt{apar\_sympl.bspl}. & Pullback/control-variate workflow. \\
$A_\parallel^h$ & Hamiltonian component of $A_\parallel$. & Hamiltonian part of parallel-potential storage. & Electromagnetic Hamiltonian and validation fields. \\
$\Lambda_\mu$ & Tensor-product B-spline basis function. & Solver B-spline basis and stencil data. & Deposition, interpolation, finite-element matrices. \\
$A_{\mu\nu}$ & Finite-element field matrix. & Solver/matrix coefficient modules. & Quasineutrality and Amp\`ere linear systems. \\
$b_\nu$ & Field right-hand side. & Source/deposition arrays such as \texttt{rho\_bspl} and \texttt{curr\_bspl}. & Charge/current deposition before field solve. \\
$(n,m)$ & Toroidal and poloidal Fourier mode numbers. & Fourier/filter arrays in field solver. & Mode filtering, Fourier solve, diagnostics. \\
\bottomrule
\end{tabular}
\caption{Important ORB5X mathematical and numerical variables. 
}
\label{tab:orb5x_variables}
\end{table}

\section{Detailed Numerical Methods}
\label{app:numerical_model}

The ORB5X time step is an operator-split PIC algorithm. The collisionless gyrokinetic characteristics and marker weights are advanced first, followed by collisions, sources, noise-control operators, profile updates, diagnostics, and electromagnetic pullback when required. The collisionless marker orbits and weight equations use a fourth-order Runge--Kutta method. ORB5X supports the standard $\delta f$ weight evolution and the direct-$\delta f$ reconstruction based on invariance of the total distribution along collisionless characteristics \cite{Allfrey2003}.

Let $g(z,t)$ be the marker distribution, chosen so that $dg/dt=0$ along collisionless trajectories. With marker weights $p_i$ and $\delta w_i$ representing $f_0$ and $\delta f$, the standard weight equations are
\begin{equation}
  g(z,t)\simeq
  \sum_{i=1}^{N_p}\frac{\delta[z-z_i(t)]}{J_z},
\label{eq:marker_distribution}
\end{equation}
and
\begin{align}
  f_0(z,t)
  &=
  P(z,t)g(z,t)
  \simeq
  \sum_{i=1}^{N_p}p_i(t)\frac{\delta[z-z_i(t)]}{J_z},\\
  \delta f(z,t)
  &=
  W(z,t)g(z,t)
  \simeq
  \sum_{i=1}^{N_p}\delta w_i(t)\frac{\delta[z-z_i(t)]}{J_z}.
\label{eq:marker_weights}
\end{align}
The weights include the physical-particle to marker-number normalization. The standard weight
equations are
\begin{equation}
  \frac{d\delta w_i}{dt}
  =
  -p_i\,\frac{1}{f_0(z_i)}\frac{df_0(z_i)}{dt},
  \qquad
  \frac{dp_i}{dt}
  =
  p_i\,\frac{1}{f_0(z_i)}\frac{df_0(z_i)}{dt}.
\label{eq:standard_weight_equations}
\end{equation}
In the direct scheme, $p_i/f_0(z_i)$ and $p_i+\delta w_i$ are used as invariants:
\begin{equation}
  p_i(t)+\delta w_i(t)=p_i(t_0)+\delta w_i(t_0),
  \qquad
  \frac{p_i(t)}{f_0(z_i(t))}
  =
  \frac{p_i(t_0)}{f_0(z_i(t_0))}.
\label{eq:direct_delta_f_invariants}
\end{equation}
ORB5X keeps these channels explicit in the particle data structures and reduced-weight routines.

Particle loading uses low-discrepancy Halton--Hammersley sequences in phase space \cite{Halton1960,Hammersley1960}. The spatial loading distribution can combine a flat component with a Gaussian radial concentration; velocity-space loading is uniform in either $|v|^2$ or $|v|^3$ up to a prescribed multiple of the thermal speed. Initial perturbations can be white-noise weights based on a van der Corput sequence \cite{Corput1935}, or mode-initialized weights over selected $(m,n)$ Fourier harmonics. After loading, the mean initial weight is subtracted. Near the magnetic axis, ORB5X follows the ORB5 practice of switching from $(s,\chi,\varphi)$ to coordinates of the form $(s\cos\chi,s\sin\chi,\varphi)$ to avoid coordinate singularities. Particles leaving the radial domain are reflected with zero perturbation weight. These details influence early-time mode growth, marker noise, and reproducibility, so they are explicit parity targets.

The field representation uses tensor-product B-splines on an $(N_s,N_\chi,N_\varphi)$ grid. For $\Psi\in\{\phi,A_\parallel\}$,
\begin{equation}
  \Psi(\bm{X},t) = \sum_\mu \Psi_\mu(t)\Lambda_\mu(\bm{X}),
  \qquad
  \Lambda_\mu=\Lambda_j^p(s)\Lambda_k^p(\chi)\Lambda_l^p(\varphi),
\label{eq:spline_field_expansion}
\end{equation}
with spline degree $p=1,2,3$. The one-dimensional cardinal B-spline basis obeys
\begin{equation}
  S_{n+1}(x) = \int S_0(x-x')S_n(x')\,dx',
  \qquad
  S_0(x)=
  \begin{cases}
  1, & |x| < 1/2,\\
  0, & \mathrm{otherwise}.
  \end{cases}
\end{equation}
Compact support, partition of unity, and derivative identities are used in interpolation, deposition, and finite-element matrix assembly.

The weak quasineutrality and Amp\`ere equations lead to linear systems
\begin{equation}
  \sum_\mu A_{\mu\nu}\Psi_\mu(t)=b_\nu(t),
\label{eq:field_linear_system}
\end{equation}
where $A_{\mu\nu}$ is the finite-element matrix for the selected model and $b_\nu$ is assembled from marker charge or current deposition. Marker quantities are deposited onto spline coefficients; the field equations are solved in the spline/Fourier basis; and fields and gradients are interpolated back to marker positions. ORB5X stores the main field and source arrays in Kokkos Views such as \texttt{rho\_bspl}, \texttt{curr\_bspl}, \texttt{phi.bspl}, \texttt{apar.bspl}, and \texttt{apar\_sympl.bspl}. Fortran-style lower bounds and guard cells are represented explicitly through offset mappings so that spline stencils and boundary semantics are preserved.

For example, in a single-species electrostatic case with adiabatic electrons and long-wavelength ion polarization, the finite-element matrix and right-hand side have the representative form
\begin{align}
  A_{\mu\nu}^{\mathrm{LWA,ad}}
  &=
  \int dV\left[
  \frac{e n_0(\psi)}{T_e(\psi)}
  \left(\Lambda_\mu\Lambda_\nu-\overline{\Lambda}_\mu\,\overline{\Lambda}_\nu\right)
  +
  \frac{n_0(\psi)m_i}{B^2}
  \nabla_\perp\Lambda_\mu\cdot\nabla_\perp\Lambda_\nu
  \right],\\
  b_\nu(t)
  &=
  \sum_{p=1}^{N}
  \frac{\delta w_p(t)}{2\pi}
  \int_0^{2\pi}d\alpha\,
  \Lambda_\nu\!\left(\bm{X}_p+\bm{\rho}_{L,p}(\alpha)\right).
\label{eq:representative_deposition}
\end{align}
The overbar denotes a flux-surface average. This example shows the two main assembly operations that ORB5X must preserve: matrix terms from model-dependent field operators and marker deposition onto gyroaveraged test functions.

The angular field solve uses discrete Fourier transforms in $(\chi,\varphi)$. Axisymmetry decouples toroidal mode numbers, reducing the solve to independent radial-poloidal systems for retained toroidal modes. The current ORB5X implementation retains the original path: poloidal transform, distributed transpose, toroidal transform, Fourier filtering, inverse toroidal transform, transpose back, and inverse poloidal transform. The source also contains local-DFT routines based on partial toroidal transforms and MPI reduction; these remain an optimization path for reducing global transposes and improving accelerator residency.

In Fourier space the system can be expressed as
\begin{equation}
  \sum_\mu \mathcal{F} A_{\mu\nu}\mathcal{F}^{-1}\,\mathcal{F}\Psi_\mu
  =
  \mathcal{F}b_\nu .
\label{eq:fourier_field_system}
\end{equation}
Axisymmetry gives $n=n'$ for the toroidal modes, so each retained toroidal mode can be solved as a radial-poloidal system,
\begin{equation}
  \sum_j\sum_m
  \widehat{\widehat{A}}_{(j,j')}^{(n,m),(n,m')}
  \widehat{\widehat{\Psi}}_j^{n,m}
  =
  \frac{\widehat{\widehat{b}}_{j'}^{n,m'}}{M^{n,p}} .
\label{eq:mode_block_system}
\end{equation}
The factor $M^{n,p}$ is the toroidal B-spline mass factor for spline degree $p$. The magnetic-axis boundary uses a unicity condition, $\Psi(s=0,\chi,\varphi,t)=\Psi(s=0,\chi=0,\varphi,t)$, while the outer radial edge normally uses Dirichlet conditions, $\Psi(s=1,\chi,\varphi,t)=0$. Annular runs use Dirichlet conditions on both radial edges. Full-FLR integral polarization variants require special treatment because the field equation is not a local differential equation.

Gyroaveraging is evaluated by sampling a Larmor ring in the poloidal plane. The code supports a fixed number of gyropoints or an adaptive number that increases for particles with larger Larmor radius \cite{Hatzky2002,Hatzky2019}. Fourier filtering is applied to deposited density and current coefficients. A rectangular filter selects an input $(n,m)$ window, and a field-aligned filter keeps poloidal modes close to $m\approx -nq(s)$, typically $m\in[-nq(s)-\Delta m,-nq(s)+\Delta m]$ \cite{Jolliet2007}. This reduces marker noise and removes modes inconsistent with the gyrokinetic ordering.

The gyropoint positions are parameterized as
\begin{equation}
  \bm{x}(\alpha)
  =
  \bm{X}
  +\rho\frac{\nabla s}{|\nabla s|}\cos\alpha
  +\rho\frac{\bm{b}\times\nabla s}{|\bm{b}\times\nabla s|}\sin\alpha .
\label{eq:gyropoint_position}
\end{equation}
The gradient of a gyroaveraged scalar field is approximated by integrating the field gradient over the ring,
\begin{equation}
  \nabla_{\bm{X}}\langle\phi_1\rangle
  =
  \frac{1}{2\pi}\oint_0^{2\pi}
  \nabla_{\bm{X}}\phi_1(\bm{X}+\bm{\rho})\,d\alpha .
\label{eq:gyroaveraged_gradient}
\end{equation}
The implementation can either retain the finite-Larmor-radius correction terms in this gradient or use the common approximation in which the gradient at the gyropoint position is used directly.

\subsection{Electromagnetic Cancellation and Pullback}

Electromagnetic gyrokinetic PIC simulations suffer from the cancellation problem:
in the parallel Amp\`ere equation, the large skin term proportional to $A_\parallel$
nearly cancels the adiabatic part of the marker current, while the physical residual
can be much smaller than either term. ORB5 mitigates this with the mixed-variable
pullback formulation \cite{Mishchenko2017,Hatzky2019,Mishchenko2018a}, which is the
form preserved in ORB5X.

The mixed-variable formulation splits the perturbed parallel vector potential into
symplectic and Hamiltonian parts,
\begin{equation}
  A_\parallel = A_\parallel^{\mathrm{s}} + A_\parallel^{\mathrm{h}} .
\label{eq:apar_split}
\end{equation}
The symplectic component modifies the phase-space structure, while the Hamiltonian
component enters the perturbed Hamiltonian and the electromagnetic field solve. In
the notation used in ORB5X, \texttt{apar\_sympl.bspl} stores
$A_\parallel^{\mathrm{s}}$, \texttt{apar.bspl} stores the active Hamiltonian
component during the Amp\`ere solve, and \texttt{apar\_ham\_bspl} keeps the
Hamiltonian component for diagnostics and verification.

With this split, the leading electromagnetic perturbation to the gyrocenter
characteristics can be written in the ORB5 mixed-variable form
\begin{align}
  \dot{\bm{X}}^{(1)}
  &=
  \frac{\bm{b}}{B_\parallel^*}\times
  \nabla\left\langle
    \phi - v_\parallel A_\parallel^{\mathrm{s}}
           - v_\parallel A_\parallel^{\mathrm{h}}
  \right\rangle
  -\frac{q_s}{m_s}\left\langle A_\parallel^{\mathrm{h}}\right\rangle
  \bm{b}_0^* ,
\label{eq:mixed_dotx}\\
  \dot{v}_\parallel^{(1)}
  &=
  -\frac{q_s}{m_s}
  \left[
    \bm{b}^*\cdot
    \nabla\left\langle \phi-v_\parallel A_\parallel^{\mathrm{h}}\right\rangle
    +\frac{\partial}{\partial t}
    \left\langle A_\parallel^{\mathrm{s}}\right\rangle
  \right]
  -\frac{\mu}{m_s}
  \frac{\bm{b}\times\nabla B}{B_\parallel^*}
  \cdot\nabla\left\langle A_\parallel^{\mathrm{s}}\right\rangle ,
\label{eq:mixed_dotv}
\end{align}
where
\begin{equation}
  \bm{b}^*
  =
  \bm{b}_0^*
  +
  \frac{\nabla\left\langle A_\parallel^{\mathrm{s}}\right\rangle\times\bm{b}}
       {B_\parallel^*},
  \qquad
  \bm{b}_0^*
  =
  \bm{b}
  +
  \frac{m_s v_\parallel}{q_s B_\parallel^*}\nabla\times\bm{b}.
\label{eq:mixed_bstar}
\end{equation}
The nonlinear magnetic-flutter term associated with
$\nabla\langle A_\parallel^{\mathrm{s}}\rangle$ is therefore part of the same
mixed-variable structure as the pullback. In the ORB5X particle pusher this appears
through the \texttt{MIXED} parallel-momentum branch, where gradients of
\texttt{apar.bspl} and \texttt{apar\_sympl.bspl} enter the electric and magnetic
flutter contributions.

For the Ohm-law pullback option, the symplectic component is advanced with
\begin{equation}
  \frac{\partial A_\parallel^{\mathrm{s}}}{\partial t}
  +
  \bm{b}\cdot\nabla\phi
  =
  0 ,
\label{eq:apar_sympl_ohm}
\end{equation}
implemented in ORB5X through \texttt{EfluidModule::apar\_ohm} and selected by
\texttt{nsel\_pullback=Ohm}. The Hamiltonian component is obtained from the
mixed-variable Amp\`ere equation,
\begin{equation}
  \sum_s \frac{\beta_s}{\rho_s^2} A_\parallel^{\mathrm{h}}
  -\nabla_\perp^2 A_\parallel^{\mathrm{h}}
  =
  \mu_0\sum_s \overline{j}_{\parallel 1s}
  +
  \nabla_\perp^2 A_\parallel^{\mathrm{s}},
\label{eq:mixed_ampere}
\end{equation}
coupled to the usual long-wavelength quasineutrality equation for $\phi$. The
source term $\nabla_\perp^2 A_\parallel^{\mathrm{s}}$ is the current correction
represented in ORB5X by \texttt{curr\_sympl\_bspl}; it must be assembled and
subtracted consistently with the deposited marker current.

At the end of a time step, the pullback redefines the split without changing the
physical distribution function:
\begin{align}
  A_{\parallel,\mathrm{new}}^{\mathrm{s}}(t_i)
  &=
  A_{\parallel,\mathrm{old}}^{\mathrm{s}}(t_i)
  +
  A_{\parallel,\mathrm{old}}^{\mathrm{h}}(t_i)
  =
  A_\parallel(t_i),
\label{eq:pullback_apar_s}\\
  A_{\parallel,\mathrm{new}}^{\mathrm{h}}(t_i) &= 0 .
\label{eq:pullback_apar_h}
\end{align}
The corresponding nonlinear pullback transforms the marker representation as
\begin{align}
  f_{1s,\mathrm{new}}
  \left(v_\parallel^{\mathrm{s}},A_\parallel^{\mathrm{s}}\right)
  &=
  f_{1s,\mathrm{old}}
  \left(v_\parallel^{\mathrm{m}},A_\parallel^{\mathrm{s}},A_\parallel^{\mathrm{h}}\right)
  +
  F_{0s}\left(v_\parallel^{\mathrm{m}}\right)
  -
  F_{0s}\left(v_\parallel^{\mathrm{s}}\right),
\label{eq:nonlinear_pullback_f}\\
  v_\parallel^{\mathrm{s}}
  &=
  v_\parallel^{\mathrm{m}}
  -
  \frac{q_s}{m_s}
  \left\langle A_{\parallel,\mathrm{old}}^{\mathrm{h}}\right\rangle .
\label{eq:nonlinear_pullback_v}
\end{align}
The linearized variant used by the \texttt{nl\_lin\_pullback} flag keeps the
particle coordinate unchanged and updates only the perturbation weight,
\begin{equation}
  f_{1s,\mathrm{new}}
  =
  f_{1s,\mathrm{old}}
  +
  \frac{q_s}{m_s}
  \left\langle A_{\parallel,\mathrm{old}}^{\mathrm{h}}\right\rangle
  \frac{\partial F_{0s}}{\partial v_\parallel}.
\label{eq:linear_pullback}
\end{equation}
In the ORB5X time loop, \texttt{ParticleModule::pullback} applies the marker
weight or coordinate transformation, and \texttt{OneStep::apply\_apar\_pullback\_poststep}
performs the field redefinition
\texttt{apar\_sympl.bspl += apar.bspl}, stages the Ohm-law field, copies the
Hamiltonian component to \texttt{apar\_ham\_bspl}, and resets \texttt{apar.bspl}.
This ordering is a high-priority Fortran-parity point because an apparently small
change in the split-field handoff changes the cancellation balance and can be
visible in $A_\parallel^{\mathrm{s}}$, $A_\parallel^{\mathrm{h}}$, and the
parallel-current residual even when electrostatic diagnostics agree.

\subsection{Noise Control}

Noise control and source terms follow the ORB5 methods. The modified Krook operator damps
non-axisymmetric fluctuations while projecting out selected conserved moments such as density,
parallel flow, zonal flow, or kinetic energy \cite{McMillan2008}. It is written
\begin{equation}
  S_K^{\mathrm{NC}}=-\gamma_K\delta f+S_K^{\mathrm{corr}},
  \qquad
  S_K^{\mathrm{corr}}=\sum_{i=1}^{N_{\mathrm{mom}}}g_i(s)M_i f_0,
\end{equation}
where the coefficients $g_i(s)$ are chosen so that
\begin{equation}
  \overline{\int dW\,M_j S_K^{\mathrm{NC}}}=0
\end{equation}
for each protected moment $M_j$. This gives a small linear system on each radial bin,
\begin{equation}
  \sum_i S_{ij}(s,t)g_i(s,t)=\delta S_j(s,t),
\end{equation}
with $S_{ij}$ obtained from flux-surface averages of $M_iM_jf_0$ and $\delta S_j$ from the moment of
$\gamma_K\delta f$.

Coarse graining smooths marker weights in field-aligned phase-space bins
\cite{Chen2007,Vernay2012}. The bin coordinates are $(s,z,\chi,\lambda,\epsilon)$ with
\begin{equation}
  z=\varphi-q(s)\left[\chi-\chi_0(\chi)\right],
\end{equation}
so the smoothing bins follow the field while remaining compatible with the toroidal domain
decomposition. The weight update is proportional to
\begin{equation}
  \mathcal{N}\Delta t\,\gamma_{\mathrm{cg}}(\overline{w}-w),
\end{equation}
where $\overline{w}$ is the bin-averaged marker weight. The quadtree method instead pairs nearby
markers in velocity space inside configuration-space bins and sets
\begin{align}
  w_1^{\mathrm{new}}&=(1-\Gamma)w_1^{\mathrm{old}}+\Gamma\overline{w},\\
  w_2^{\mathrm{new}}&=(1-\Gamma)w_2^{\mathrm{old}}+\Gamma\overline{w},\\
  \Gamma&=\exp\left[-\frac{(v_1^x-v_2^x)^2+(v_1^y-v_2^y)^2}{h_v^2}\right],
  \qquad
  \overline{w}=\frac{w_1^{\mathrm{old}}+w_2^{\mathrm{old}}}{2}.
\end{align}
This operation conserves total marker weight by construction \cite{Sonnendrucker2015}.

Thermal-relaxation heating keeps the energy-dependent perturbation close to the initial distribution
while subtracting the flux-surface-averaged density response, so it does not act as an artificial
charge source \cite{McMillan2008,McMillan2014}:
\begin{equation}
  S_{H1}
  =
  -\gamma_H
  \left[
    \delta f(\epsilon,s)
    -
    f_0(\epsilon,s)
    \frac{\overline{\delta f(\epsilon,s)}}{\overline{f_0(\epsilon,s)}}
  \right].
\end{equation}
A fixed-power source uses
\begin{equation}
  S_{H2}=\gamma_R(s)\frac{\partial f_0}{\partial T},
\end{equation}
and can be paired with an edge sink or a boundary Krook region.

Conservation diagnostics are a central end-to-end check of the implementation. The variational formulation implies an energy balance in which the time derivative of kinetic energy, evaluated from particle characteristics, is balanced by the time derivative of field energy, evaluated from grid quantities \cite{Tronko2016}. ORB5X translates the field-energy, $j\cdot E$, entropy-production, radial-flux, moment, phase-space, and three-dimensional HDF5 diagnostic infrastructure. These diagnostics couple particle push, deposition, field solve, interpolation, normalization, MPI reductions, and output layout, making them stronger validation targets than isolated kernels.

\subsection{Overall Time-Step Algorithm}
\label{app:overall_algorithm}

The outer ORB5X loop advances a coupled particle--field system. At the beginning of a time step the particle arrays contain marker positions, canonical momenta, magnetic moments, statistical weights, and species-dependent auxiliary variables; the field arrays contain the B-spline/Fourier coefficients of $\phi$ and, for electromagnetic simulations, the split parallel vector potential $A_\parallel=A_\parallel^s+A_\parallel^h$. A single step from $t^n$ to $t^{n+1}=t^n+\Delta t$ can be viewed as the following sequence.

\begin{algorithm}[H]
\footnotesize
\DontPrintSemicolon
\caption{One ORB5X time step for nonlinear electromagnetic gyrokinetic PIC}
\label{alg:orb5x_timestep}
\KwIn{Marker state at $t^n$, field coefficients at $t^n$, profiles, sources, collision operators, MPI topology.}
\KwOut{Marker state, field coefficients, diagnostics, and migrated particles at $t^{n+1}$.}

\If{electromagnetic split representation is active}{
  prepare the pullback state and the current $A_\parallel^s,A_\parallel^h$ representation associated with the canonical momentum definition \eqref{eq:pz_definition}\;
}

\For{$k=1,\ldots,4$ Runge--Kutta stages}{
  construct stage marker coordinates and weights from the RK accumulator, using the marker representation \eqref{eq:marker_distribution}--\eqref{eq:marker_weights}\;

  deposit particle moments to the grid:
  charge density, polarization sources, parallel current, and species moments required by the selected model; representative spline deposition is given by \eqref{eq:representative_deposition}\;

  reduce clone-local deposits and exchange/update subdomain guard data\;

  assemble the finite-element right-hand sides in the B-spline basis \eqref{eq:spline_field_expansion} and apply Fourier/mode filters\;

  solve the field equations:
  quasineutrality for $\phi$ and, in electromagnetic runs, Amp\`ere's equation for $A_\parallel^h$ using the linear systems \eqref{eq:field_linear_system}, \eqref{eq:fourier_field_system}, and \eqref{eq:mode_block_system}\;

  compute field derivatives and gyroaveraged quantities using the gyropoint geometry and gradient approximation \eqref{eq:gyropoint_position}--\eqref{eq:gyroaveraged_gradient}\;

  interpolate fields, gradients, gyroaverages, and magnetic-geometry factors to marker positions for the Hamiltonian coupling \eqref{eq:first_order_hamiltonian}\;

  evaluate the gyrocenter characteristics \eqref{eq:hamiltonian_characteristics} or the electromagnetic form \eqref{eq:em_characteristics}, and the weight right-hand sides \eqref{eq:standard_weight_equations}\;

  accumulate the RK increment for marker positions, $p_z$, and marker weights\;
}

update marker coordinates and weights with the fourth-order RK combination\;

\If{electromagnetic pullback is active}{
  update the split $A_\parallel^s/A_\parallel^h$ representation and transform marker weights or coordinates consistently with the mixed-variable pullback relations \eqref{eq:pullback_apar_s}--\eqref{eq:linear_pullback}\;
}

apply collisions, heating/source terms, Krook or quadtree noise-control operators, and profile updates as requested by the input model\;

migrate particles that crossed toroidal subdomain boundaries and fill local particle-array holes\;

compute diagnostics, conservation balances, moments, fluxes, and HDF5 output fields\;
\end{algorithm}

The Runge--Kutta loop is written above as if all particle and field quantities were recomputed at each stage. This is the algorithmic requirement for nonlinear simulations: particle deposition depends on the stage particle state, the field solve depends on the deposited charge and current, and the force on each particle depends on the freshly solved and interpolated stage fields. Linearized runs use the same data path but selectively suppress nonlinear feedback terms in the characteristics, weight equations, or deposited sources according to the input model. This common ordering is useful for validation because electrostatic, linear electromagnetic, and nonlinear electromagnetic simulations differ mainly in which terms are active, not in the top-level particle--field handshake.

For electromagnetic simulations, the sensitive part of the time step is the treatment of $A_\parallel$. ORB5X follows the ORB5 split-field workflow in which the symplectic component $A_\parallel^s$ enters the phase-space structure and the Hamiltonian component $A_\parallel^h$ enters the Hamiltonian perturbation and the Amp\`ere solve. During a nonlinear electromagnetic RK stage, marker currents and weights are deposited, the parallel-vector-potential equation is solved consistently with the current split representation, and the resulting $A_\parallel$ and its derivatives are interpolated back to the markers for the next force evaluation. The pullback step transfers information between $A_\parallel^s$, $A_\parallel^h$, and the marker weights so that the numerical representation changes without changing the represented distribution function. Consequently, the time-loop validation must compare not only final fields but also the ordering of deposition, field solve, interpolation, pullback, and diagnostics.

In parallel runs the same logical sequence is executed on every MPI rank. Depositions are local to the rank's clone and toroidal subdomain, then reduced over clone communicators and exchanged over subdomain communicators before the field solve. Particle migration is delayed until after the particle push so that each RK stage operates on a consistent local marker set. On accelerator backends the particle loops, deposition kernels, interpolation, and local algebra are expressed with Kokkos where possible, while FFTW, LAPACK, MPI, and HDF5 paths may stage through host-accessible buffers.

\normalsize

\section{LLM-Assisted Source Translation Workflow}
\label{app:llm_translation}

The ORB5X translation was assisted by prompt engineering and specialized large language models
(LLMs). The LLMs were not used as an autonomous replacement for numerical verification. Instead,
they were used as structured translation assistants: a Fortran routine, its dependent types, and the
relevant neighboring routines were supplied as context; the model was asked to produce a conservative
C++17/Kokkos translation; and the output was then checked against the original source, unit tests,
and full-code benchmark data. This appendix records the style of prompt used in the translation
because it is part of the practical methodology of moving a mature scientific Fortran code to a
portable C++ implementation.

\begin{figure}[p]
\centering
\begin{minipage}{0.98\linewidth}
\scriptsize
\begin{lstlisting}[basicstyle=\ttfamily\tiny,breaklines=true,columns=fullflexible,frame=single,aboveskip=2pt,belowskip=2pt]
Task: Run an implementation campaign translating the attached ORB5 Fortran
routine to C++17/Kokkos for ORB5X.
Context:
- The goal is numerical parity with ORB5, not redesign.
- Preserve the mathematical algorithm, update order, boundary conditions,
  reductions, and diagnostic side effects.
- Use Kokkos::View for performance-critical arrays and Kokkos kernels for
  particle/grid loops when this does not change the algorithm.
- Keep MPI communication semantics identical unless explicitly asked otherwise.
- Treat each patch as provisional until it improves or preserves HDF5 parity.
Inputs:
1. Fortran source routine:
   <paste routine here>
2. Dependent Fortran module declarations and derived types:
   <paste relevant declarations here>
3. Current ORB5X C++ headers/classes for this module:
   <paste relevant C++ context here>
4. Existing parity tests or reference output:
   <paste expected formulas, array bounds, or diagnostics here>
Build and run commands:
- ORB5X:
  <build command, e.g. cmake --build build -j>
  <run command, e.g. mpirun -np P build/orb5x input>
- Fortran ORB5 reference:
  <build command and matching reference run command>
- HDF5 comparison:
  <compare orb5x/output.h5 against orb5/reference.h5>
Translation rules:
- Preserve Fortran array lower and upper bounds. If a Fortran array is indexed
  as a(i0:i1), create an explicit offset mapping in C++ rather than silently
  assuming zero-based indexing.
- Mark every intentional index shift in a short comment.
- Preserve loop inclusivity. Fortran "do i=a,b" includes b.
- Preserve column-major mathematical ordering. If storage layout changes, separate
  physical indices from memory indices using helper accessors.
- Preserve real kinds and integer widths. Map ORB5 R4/I4/I8/RKIND to the
  corresponding ORB5X aliases, not raw C++ primitive types.
- Preserve reductions and identify deterministic or MPI-reduced sums.
- Preserve guard-cell, periodic, and radial-boundary behavior exactly.
- Preserve all abort/error conditions, even if they look redundant.
- Do not simplify formulas, reorder floating-point operations, or fuse loops unless
  the change is explicitly called out as a proposed optimization.
Campaign loop:
1. Implement the smallest coherent patch.
2. Build ORB5X and the Fortran reference if needed.
3. Run the same benchmark/input script with both codes.
4. Compare output HDF5 datasets, shapes, attributes, and selected norms.
5. Accept the patch only if the number of discrepancies or the numerical error is
   reduced, or if it is unchanged while enabling the next required porting step.
6. If parity worsens, explain the suspected cause and revise or revert the patch.
Output format for each campaign iteration:
1. Mathematical role and array mapping table with offsets.
2. Patch summary and files changed.
3. Build/run commands executed.
4. HDF5 comparison: changed datasets, norms, discrepancy count.
5. Accept/reject decision and next implementation step.
\end{lstlisting}
\end{minipage}
\caption{Representative prompt structure used when translating ORB5 Fortran kernels to ORB5X C++/Kokkos.}
\label{fig:translation_prompt}
\end{figure}

Several recurring issues required explicit prompt constraints and human review. The most common was
indexing. Fortran arrays in ORB5 frequently use one-based indexing, custom lower bounds, and guard
cell ranges such as $0:N+1$ or $-p:N+p$. A direct C++ rewrite can compile while being wrong by one
cell. ORB5X therefore uses explicit offset rules when preserving Fortran semantics, especially in
spline stencils, Fourier-mode windows, particle buffers, and guard-cell updates. A related issue is
loop inclusivity: Fortran upper bounds are included, while C++ loop conditions must be written
deliberately to preserve the same iteration set.

Memory layout was another source of risk. Fortran stores multidimensional arrays in column-major
order and ORB5 often encodes numerical meaning in the order of indices. ORB5X separates the
mathematical index order from the Kokkos memory layout whenever needed, so that performance tuning
does not obscure the finite-element or particle-field mapping. Similar care is required for
floating-point reductions: changing the order of summation can alter particle deposition,
diagnostic totals, and MPI-reduced field quantities. For this reason, the LLM-generated translations
were treated as first drafts and were checked with small deterministic tests before being trusted in
full benchmark runs.

The prompt also asked the model to report uncertainty rather than silently filling gaps. This was
important for code paths involving MPI datatypes, HDF5 output, host/device staging, and mixed
precision variables. In practice, the most useful LLM output was not only the translated C++ code,
but also the accompanying audit list: which arrays required offset mappings, which reductions were
semantically important, which boundary conditions were implicit in the Fortran loop bounds, and
which tests would expose a likely parity error. Later prompts also made the acceptance criterion
explicit: each implementation campaign had to compile and run ORB5X and, when practical, the
Fortran reference on the same input; the resulting HDF5 files were compared dataset-by-dataset; and
a patch was accepted only if the number of discrepancies or the measured error decreased, or if it
remained unchanged while enabling the next necessary translation step. This turned the LLM workflow
from a one-shot translation task into an iterative numerical-convergence process.

\section{Additional performance portability results}
\label{sec:additional-portability-results}

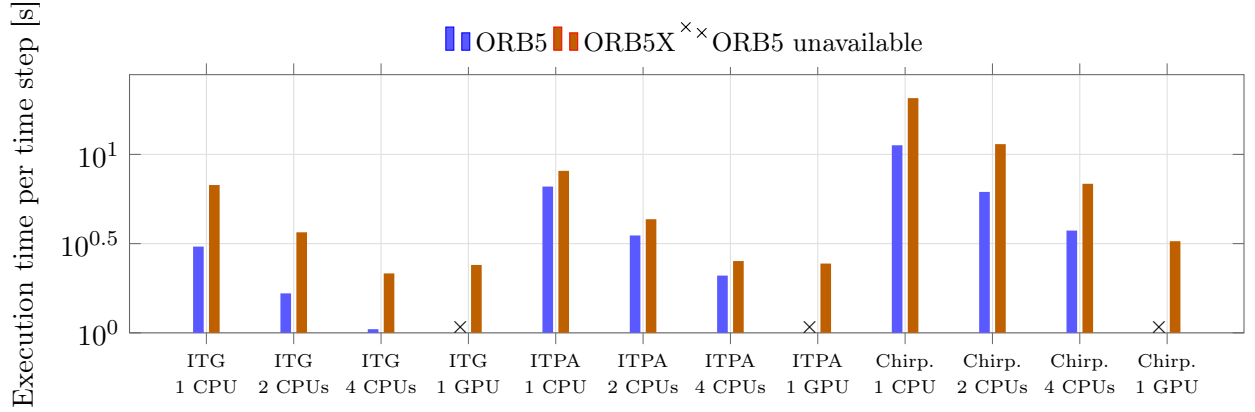
\begin{figure}
\centering
\begin{tikzpicture}
\begin{axis}[
    ybar,
    width=0.98\linewidth,
    height=\performancechartheight,
    bar width=4pt,
    ymode=log,
    log basis y=10,
    ymin=1,
    ylabel={Execution time per time step [s]},
    symbolic x coords={
        ITG CPU 1,
        ITG CPU 2,
        ITG CPU 4,
        ITG CUDA 1,
        ITPA CPU 1,
        ITPA CPU 2,
        ITPA CPU 4,
        ITPA CUDA 1,
        Chirping CPU 1,
        Chirping CPU 2,
        Chirping CPU 4,
        Chirping CUDA 1
    },
    xtick={
        ITG CPU 1,
        ITG CPU 2,
        ITG CPU 4,
        ITG CUDA 1,
        ITPA CPU 1,
        ITPA CPU 2,
        ITPA CPU 4,
        ITPA CUDA 1,
        Chirping CPU 1,
        Chirping CPU 2,
        Chirping CPU 4,
        Chirping CUDA 1
    },
    xticklabels={
        {ITG\\1 CPU},
        {ITG\\2 CPUs},
        {ITG\\4 CPUs},
        {ITG\\1 GPU},
        {ITPA\\1 CPU},
        {ITPA\\2 CPUs},
        {ITPA\\4 CPUs},
        {ITPA\\1 GPU},
        {Chirp.\\1 CPU},
        {Chirp.\\2 CPUs},
        {Chirp.\\4 CPUs},
        {Chirp.\\1 GPU}
    },
    xticklabel style={align=center,font=\tiny},
    enlarge x limits=0.08,
    legend style={
        at={(0.5,1.03)},
        anchor=south,
        legend columns=3,
        draw=none,
        font=\small
    },
    grid=major,
    major grid style={gray!25},
    axis line style={black!60},
    tick style={black!60},
]
\addplot+[draw=none, fill=blue!65] coordinates {
    (ITG CPU 1,3.04390120)
    (ITG CPU 2,1.66435730)
    (ITG CPU 4,1.04830000)
    (ITPA CPU 1,6.60366290)
    (ITPA CPU 2,3.51375020)
    (ITPA CPU 4,2.09351640)
    (Chirping CPU 1,11.25099800)
    (Chirping CPU 2,6.16321930)
    (Chirping CPU 4,3.74229960)
};
\addlegendentry{ORB5}
\addplot+[draw=none, fill=orange!75!black] coordinates {
    (ITG CPU 1,6.73593550)
    (ITG CPU 2,3.66028770)
    (ITG CPU 4,2.15373570)
    (ITG CUDA 1,2.40145520)
    (ITPA CPU 1,8.08646200)
    (ITPA CPU 2,4.33268190)
    (ITPA CPU 4,2.52545570)
    (ITPA CUDA 1,2.44488410)
    (Chirping CPU 1,20.67166000)
    (Chirping CPU 2,11.41642700)
    (Chirping CPU 4,6.84561900)
    (Chirping CUDA 1,3.26277410)
};
\addlegendentry{ORB5X}
\node[font=\footnotesize, xshift=-3pt] at (axis cs:ITG CUDA 1,1.08) {$\times$};
\node[font=\footnotesize, xshift=-3pt] at (axis cs:ITPA CUDA 1,1.08) {$\times$};
\node[font=\footnotesize, xshift=-3pt] at (axis cs:Chirping CUDA 1,1.08) {$\times$};
\addlegendimage{only marks, mark=x, mark size=2.5pt, black}
\addlegendentry{ORB5 unavailable}
\end{axis}
\end{tikzpicture}
\caption{Performance of ORB5X versus ORB5 on Legion Pro 5i laptop with Intel Core i9-14900HX CPU and NVIDIA GeForce RTX 5070 GPU for representative light validation ITG, ITPA, and Chirping test cases.
}
\label{fig:performance_bar_chart_laptop}
\end{figure}

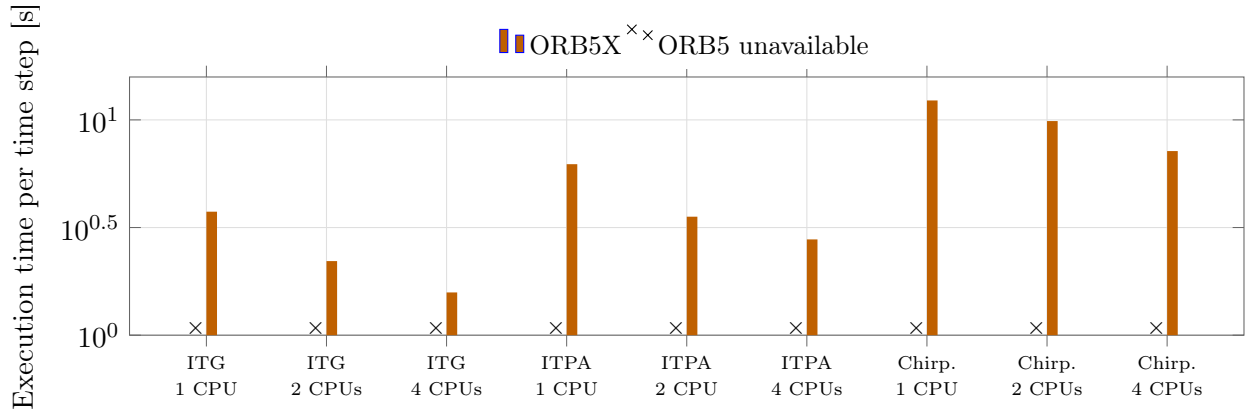
\begin{figure}
\centering
\begin{tikzpicture}
\begin{axis}[
    ybar,
    width=0.98\linewidth,
    height=\performancechartheight,
    bar width=4pt,
    ymode=log,
    log basis y=10,
    ymin=1,
    ylabel={Execution time per time step [s]},
    symbolic x coords={
        ITG CPU 1,
        ITG CPU 2,
        ITG CPU 4,
        ITPA CPU 1,
        ITPA CPU 2,
        ITPA CPU 4,
        Chirping CPU 1,
        Chirping CPU 2,
        Chirping CPU 4
    },
    xtick={
        ITG CPU 1,
        ITG CPU 2,
        ITG CPU 4,
        ITPA CPU 1,
        ITPA CPU 2,
        ITPA CPU 4,
        Chirping CPU 1,
        Chirping CPU 2,
        Chirping CPU 4
    },
    xticklabels={
        {ITG\\1 CPU},
        {ITG\\2 CPUs},
        {ITG\\4 CPUs},
        {ITPA\\1 CPU},
        {ITPA\\2 CPU},
        {ITPA\\4 CPUs},
        {Chirp.\\1 CPU},
        {Chirp.\\2 CPUs},
        {Chirp.\\4 CPUs}
    },
    xticklabel style={align=center,font=\tiny},
    enlarge x limits=0.08,
    legend style={
        at={(0.5,1.03)},
        anchor=south,
        legend columns=2,
        draw=none,
        font=\small
    },
    grid=major,
    major grid style={gray!25},
    axis line style={black!60},
    tick style={black!60},
]
\addplot+[draw=none, fill=orange!75!black, bar shift=2pt] coordinates {
    (ITG CPU 1,3.74668033)
    (ITG CPU 2,2.20879267)
    (ITG CPU 4,1.57955267)
    (ITPA CPU 1,6.22689767)
    (ITPA CPU 2,3.55123300)
    (ITPA CPU 4,2.78499000)
    (Chirping CPU 1,12.31044909)
    (Chirping CPU 2,9.87575000)
    (Chirping CPU 4,7.16521400)
};
\addlegendentry{ORB5X}
\node[font=\footnotesize, xshift=-4pt] at (axis cs:ITG CPU 1,1.08) {$\times$};
\node[font=\footnotesize, xshift=-4pt] at (axis cs:ITG CPU 2,1.08) {$\times$};
\node[font=\footnotesize, xshift=-4pt] at (axis cs:ITG CPU 4,1.08) {$\times$};
\node[font=\footnotesize, xshift=-4pt] at (axis cs:ITPA CPU 1,1.08) {$\times$};
\node[font=\footnotesize, xshift=-4pt] at (axis cs:ITPA CPU 2,1.08) {$\times$};
\node[font=\footnotesize, xshift=-4pt] at (axis cs:ITPA CPU 4,1.08) {$\times$};
\node[font=\footnotesize, xshift=-4pt] at (axis cs:Chirping CPU 1,1.08) {$\times$};
\node[font=\footnotesize, xshift=-4pt] at (axis cs:Chirping CPU 2,1.08) {$\times$};
\node[font=\footnotesize, xshift=-4pt] at (axis cs:Chirping CPU 4,1.08) {$\times$};
\addlegendimage{only marks, mark=x, mark size=2.5pt, black}
\addlegendentry{ORB5 unavailable}
\end{axis}
\end{tikzpicture}
\caption{Performance of ORB5X
on Macbook Air laptop with Apple M4 chip for representative light validation ITG, ITPA, and Chirping test cases. 
}
\label{fig:performance_bar_chart_mac_laptop}
\end{figure}

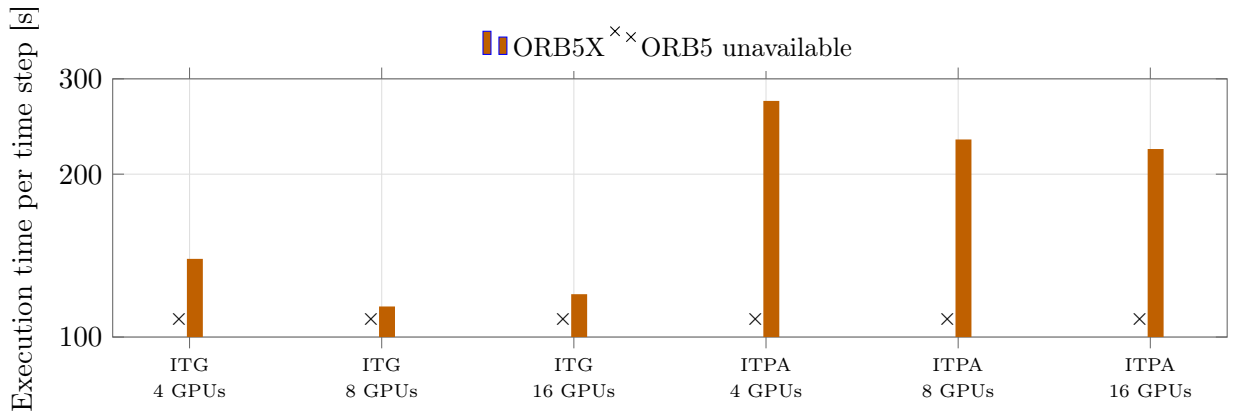
\begin{figure}
\centering
\begin{tikzpicture}
\begin{axis}[
    ybar,
    width=0.98\linewidth,
    height=\performancechartheight,
    bar width=6pt,
    ymode=log,
    log basis y=10,
    log ticks with fixed point,
    log origin y=infty,
    ymin=100,
    ymax=300,
    ytick={100,200,300},
    ylabel={Execution time per time step [s]},
    symbolic x coords={
        ITG GPU 4,
        ITG GPU 8,
        ITG GPU 16,
        ITPA GPU 4,
        ITPA GPU 8,
        ITPA GPU 16
    },
    xtick=data,
    xticklabels={
        {ITG\\4 GPUs},
        {ITG\\8 GPUs},
        {ITG\\16 GPUs},
        {ITPA\\4 GPUs},
        {ITPA\\8 GPUs},
        {ITPA\\16 GPUs}
    },
    xticklabel style={align=center,font=\tiny},
    enlarge x limits=0.08,
    legend style={
        at={(0.5,1.03)},
        anchor=south,
        legend columns=2,
        draw=none,
        font=\small
    },
    grid=major,
    major grid style={gray!25},
    axis line style={black!60},
    tick style={black!60},
]
\addplot+[draw=none, fill=orange!75!black, bar shift=2pt] coordinates {
    (ITG GPU 4,139.466000)
    (ITG GPU 8,113.891010)
    (ITG GPU 16,119.997500)
    (ITPA GPU 4,273.113045)
    (ITPA GPU 8,231.784000)
    (ITPA GPU 16,222.563360)
};
\addlegendentry{ORB5X}
\node[font=\footnotesize, xshift=-4pt] at (axis cs:ITG GPU 4,108) {$\times$};
\node[font=\footnotesize, xshift=-4pt] at (axis cs:ITG GPU 8,108) {$\times$};
\node[font=\footnotesize, xshift=-4pt] at (axis cs:ITG GPU 16,108) {$\times$};
\node[font=\footnotesize, xshift=-4pt] at (axis cs:ITPA GPU 4,108) {$\times$};
\node[font=\footnotesize, xshift=-4pt] at (axis cs:ITPA GPU 8,108) {$\times$};
\node[font=\footnotesize, xshift=-4pt] at (axis cs:ITPA GPU 16,108) {$\times$};
\addlegendimage{only marks, mark=x, mark size=2.5pt, black}
\addlegendentry{ORB5 unavailable}
\end{axis}
\end{tikzpicture}
\caption{
Performance of ORB5X 
on Pitagora at CINECA with NVIDIA H100 GPUs for the heavy validation ITG and ITPA test cases. 
}
\label{fig:performance_bar_chart_pitagora_h100}
\end{figure}

\begin{figure}
\centering
\begin{tikzpicture}
\begin{axis}[
    ybar,
    width=0.98\linewidth,
    height=\performancechartheight,
    bar width=6pt,
    ymode=log,
    log basis y=10,
    log ticks with fixed point,
    log origin y=infty,
    ymin=10,
    ymax=100,
    ytick={10,100},
    ylabel={Execution time per time step [s]},
    symbolic x coords={
        ITG GPU 4,
        ITG GPU 8,
        ITG GPU 16,
        ITPA GPU 4,
        ITPA GPU 8,
        ITPA GPU 16,
        Chirping GPU 4,
        Chirping GPU 8,
        Chirping GPU 16
    },
    xtick=data,
    xticklabels={
        {ITG\\4 GPUs},
        {ITG\\8 GPUs},
        {ITG\\16 GPUs},
        {ITPA\\4 GPUs},
        {ITPA\\8 GPUs},
        {ITPA\\16 GPUs},
        {Chirp.\\4 GPUs},
        {Chirp.\\8 GPUs},
        {Chirp.\\16 GPUs}
    },
    xticklabel style={align=center,font=\tiny},
    enlarge x limits=0.08,
    legend style={
        at={(0.5,1.03)},
        anchor=south,
        legend columns=2,
        draw=none,
        font=\small
    },
    grid=major,
    major grid style={gray!25},
    axis line style={black!60},
    tick style={black!60},
]
\addplot+[draw=none, fill=orange!75!black, bar shift=2pt] coordinates {
    (ITG GPU 4,39.83)
    (ITG GPU 8,30.96)
    (ITG GPU 16,25.43)
    (ITPA GPU 4,42.07)
    (ITPA GPU 8,37.21)
    (ITPA GPU 16,37.18)
    (Chirping GPU 4,55.96)
    (Chirping GPU 8,39.08)
    (Chirping GPU 16,31.27)
};
\addlegendentry{ORB5X}
\node[font=\footnotesize, xshift=-4pt] at (axis cs:ITG GPU 4,10.8) {$\times$};
\node[font=\footnotesize, xshift=-4pt] at (axis cs:ITG GPU 8,10.8) {$\times$};
\node[font=\footnotesize, xshift=-4pt] at (axis cs:ITG GPU 16,10.8) {$\times$};
\node[font=\footnotesize, xshift=-4pt] at (axis cs:ITPA GPU 4,10.8) {$\times$};
\node[font=\footnotesize, xshift=-4pt] at (axis cs:ITPA GPU 8,10.8) {$\times$};
\node[font=\footnotesize, xshift=-4pt] at (axis cs:ITPA GPU 16,10.8) {$\times$};
\node[font=\footnotesize, xshift=-4pt] at (axis cs:Chirping GPU 4,10.8) {$\times$};
\node[font=\footnotesize, xshift=-4pt] at (axis cs:Chirping GPU 8,10.8) {$\times$};
\node[font=\footnotesize, xshift=-4pt] at (axis cs:Chirping GPU 16,10.8) {$\times$};
\addlegendimage{only marks, mark=x, mark size=2.5pt, black}
\addlegendentry{ORB5 unavailable}
\end{axis}
\end{tikzpicture}
\caption{Performance of ORB5X 
on CSCS-Daint of ALPS with NVIDIA GH200 GPUs
for the heavy validation ITG, ITPA, and Chirping test cases.
 }
\label{fig:performance_bar_chart_gh200_daint}
\end{figure}
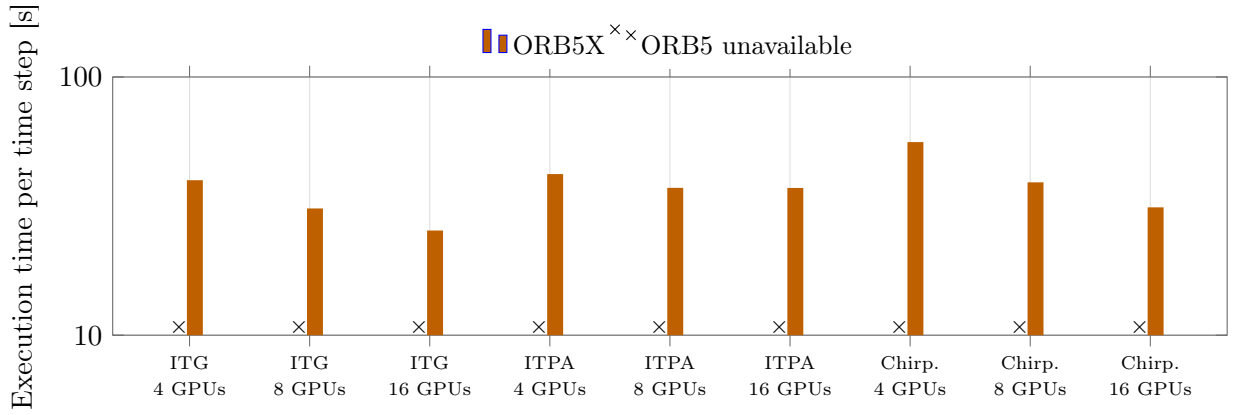

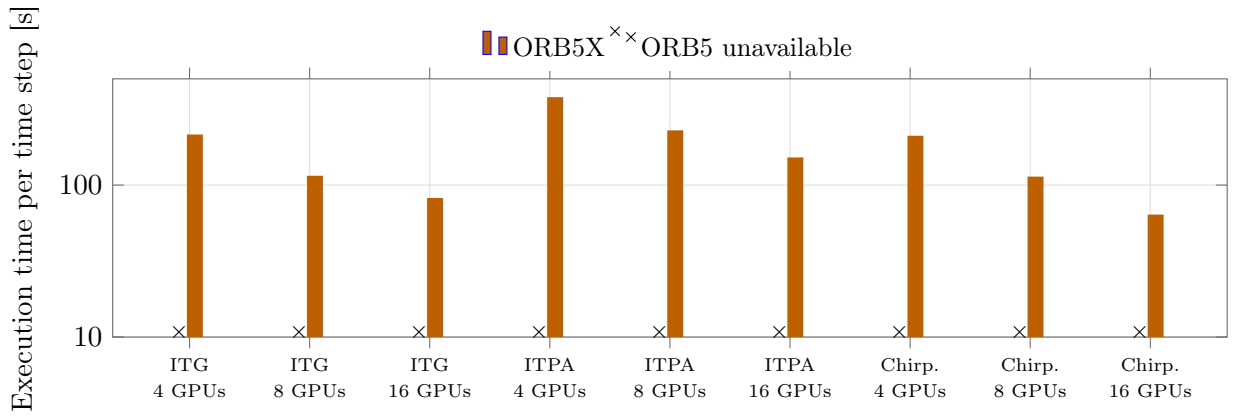
\begin{figure}
\centering
\begin{tikzpicture}
\begin{axis}[
    ybar,
    width=0.98\linewidth,
    height=\performancechartheight,
    bar width=6pt,
    ymode=log,
    log basis y=10,
    log ticks with fixed point,
    log origin y=infty,
    ymin=10,
    ymax=500,
    ytick={10,100},
    ylabel={Execution time per time step [s]},
    symbolic x coords={
        ITG GPU 4,
        ITG GPU 8,
        ITG GPU 16,
        ITPA GPU 4,
        ITPA GPU 8,
        ITPA GPU 16,
        Chirping GPU 4,
        Chirping GPU 8,
        Chirping GPU 16
    },
    xtick=data,
    xticklabels={
        {ITG\\4 GPUs},
        {ITG\\8 GPUs},
        {ITG\\16 GPUs},
        {ITPA\\4 GPUs},
        {ITPA\\8 GPUs},
        {ITPA\\16 GPUs},
        {Chirp.\\4 GPUs},
        {Chirp.\\8 GPUs},
        {Chirp.\\16 GPUs}
    },
    xticklabel style={align=center,font=\tiny},
    enlarge x limits=0.08,
    legend style={
        at={(0.5,1.03)},
        anchor=south,
        legend columns=2,
        draw=none,
        font=\small
    },
    grid=major,
    major grid style={gray!25},
    axis line style={black!60},
    tick style={black!60},
]
\addplot+[draw=none, fill=orange!75!black, bar shift=2pt] coordinates {
    (ITG GPU 4,215.148)
    (ITG GPU 8,115.148)
    (ITG GPU 16,82.324)
    (ITPA GPU 4,378.823)
    (ITPA GPU 8,229.147)
    (ITPA GPU 16,152.032)
    (Chirping GPU 4,211.149)
    (Chirping GPU 8,113.692)
    (Chirping GPU 16,63.966)
};
\addlegendentry{ORB5X}
\node[font=\footnotesize, xshift=-4pt] at (axis cs:ITG GPU 4,10.8) {$\times$};
\node[font=\footnotesize, xshift=-4pt] at (axis cs:ITG GPU 8,10.8) {$\times$};
\node[font=\footnotesize, xshift=-4pt] at (axis cs:ITG GPU 16,10.8) {$\times$};
\node[font=\footnotesize, xshift=-4pt] at (axis cs:ITPA GPU 4,10.8) {$\times$};
\node[font=\footnotesize, xshift=-4pt] at (axis cs:ITPA GPU 8,10.8) {$\times$};
\node[font=\footnotesize, xshift=-4pt] at (axis cs:ITPA GPU 16,10.8) {$\times$};
\node[font=\footnotesize, xshift=-4pt] at (axis cs:Chirping GPU 4,10.8) {$\times$};
\node[font=\footnotesize, xshift=-4pt] at (axis cs:Chirping GPU 8,10.8) {$\times$};
\node[font=\footnotesize, xshift=-4pt] at (axis cs:Chirping GPU 16,10.8) {$\times$};
\addlegendimage{only marks, mark=x, mark size=2.5pt, black}
\addlegendentry{ORB5 unavailable}
\end{axis}
\end{tikzpicture}
\caption{Performance of ORB5X 
on LUMI-G with AMD MI250x GPUs
for the heavy validation ITG, ITPA, and Chirping test cases.
 }
\label{fig:performance_bar_chart_amd_lumig}
\end{figure}

In order to demonstrate the portability of ORB5X, here we report the execution time of ORB5X for
 ITG, ITPA, and Chirping test cases on several hardware. We test the light version of these tests with affordable number of particles and mesh size for personal laptops such as Legion Pro 5i with Intel i9-14900HX CPU 
 and GeForce RTX 5070 GPU in Fig.~\ref{fig:performance_bar_chart_laptop} and 
 Apple MacBook Air with M4 chip in Fig.~\ref{fig:performance_bar_chart_mac_laptop}. In computer clusters, we test the fine simulation settings described in previous sections on NVIDIA-H100 GPUs of Pitagora at CINECA\footnote{\url{https://www.cineca.it}} in Fig.~\ref{fig:performance_bar_chart_pitagora_h100}, NVIDIA-DGX H200 GPUs of Discoverer+ at Discvoerer Super Computer\footnote{\url{https://discoverer.bg}} in Fig.~\ref{fig:performance_bar_chart_h100_discoverer}, AMD MI250x GPUs at LUMI\footnote{\url{https://lumi-supercomputer.eu}} in Fig. \ref{fig:performance_bar_chart_amd_lumig}, and NVIDIA-GH200 GPUs of CSCS-Daint ALPS\footnote{\url{https://www.cscs.ch}} in Fig.~\ref{fig:performance_bar_chart_gh200_daint} as modern HPC clusters. Here, we include ORB5 timings as a reference only in cases where the original Fortran code compiles and runs successfully on the target machine. Although the Fortran code remains faster than ORB5X as it has been fine tuned for performance over the last decade, ORB5's inability to be easily deployed on different machines reconfirms the performance portability motivation behind building ORB5X.

\begin{figure}
\centering
\begin{tikzpicture}
\begin{axis}[
    ybar,
    width=0.98\linewidth,
    height=\performancechartheight,
    bar width=6pt,
    ymode=log,
    log basis y=10,
    log ticks with fixed point,
    log origin y=infty,
    ymin=50,
    ymax=200,
    ytick={50,100,200},
    ylabel={Execution time per time step [s]},
    symbolic x coords={
        ITG GPU 4,
        ITG GPU 8,
        ITPA GPU 4,
        ITPA GPU 8,
        Chirping GPU 4,
        Chirping GPU 8,
    },
    xtick=data,
    xticklabels={
        {ITG\\4 GPUs},
        {ITG\\8 GPUs},
        {ITPA\\4 GPUs},
        {ITPA\\8 GPUs},
        {Chirp.\\4 GPUs},
        {Chirp.\\8 GPUs},
    },
    xticklabel style={align=center,font=\tiny},
    enlarge x limits=0.08,
    legend style={
        at={(0.5,1.03)},
        anchor=south,
        legend columns=2,
        draw=none,
        font=\small
    },
    grid=major,
    major grid style={gray!25},
    axis line style={black!60},
    tick style={black!60},
]
\addplot+[draw=none, fill=orange!75!black, bar shift=2pt] coordinates {
    (ITG GPU 4,99.83)
    (ITG GPU 8,87.96)
    (ITPA GPU 4,138.07)
    (ITPA GPU 8,133.21)
    (Chirping GPU 4,100.96)
    (Chirping GPU 8,62.08)
};
\addlegendentry{ORB5X}
\node[font=\footnotesize, xshift=-4pt] at (axis cs:ITG GPU 4,10.8) {$\times$};
\node[font=\footnotesize, xshift=-4pt] at (axis cs:ITG GPU 8,10.8) {$\times$};
\node[font=\footnotesize, xshift=-4pt] at (axis cs:ITPA GPU 4,10.8) {$\times$};
\node[font=\footnotesize, xshift=-4pt] at (axis cs:ITPA GPU 8,10.8) {$\times$};
\node[font=\footnotesize, xshift=-4pt] at (axis cs:Chirping GPU 4,10.8) {$\times$};
\node[font=\footnotesize, xshift=-4pt] at (axis cs:Chirping GPU 8,10.8) {$\times$};
\addlegendimage{only marks, mark=x, mark size=2.5pt, black}
\addlegendentry{ORB5 unavailable}
\end{axis}
\end{tikzpicture}
\caption{Performance of ORB5X 
on 
Discoverer+ of Discoverer Supercomputer
 with NVIDIA DGX H200 GPUs
for the heavy validation ITG, ITPA, and Chirping test cases.
 }
\label{fig:performance_bar_chart_h100_discoverer}
\end{figure}
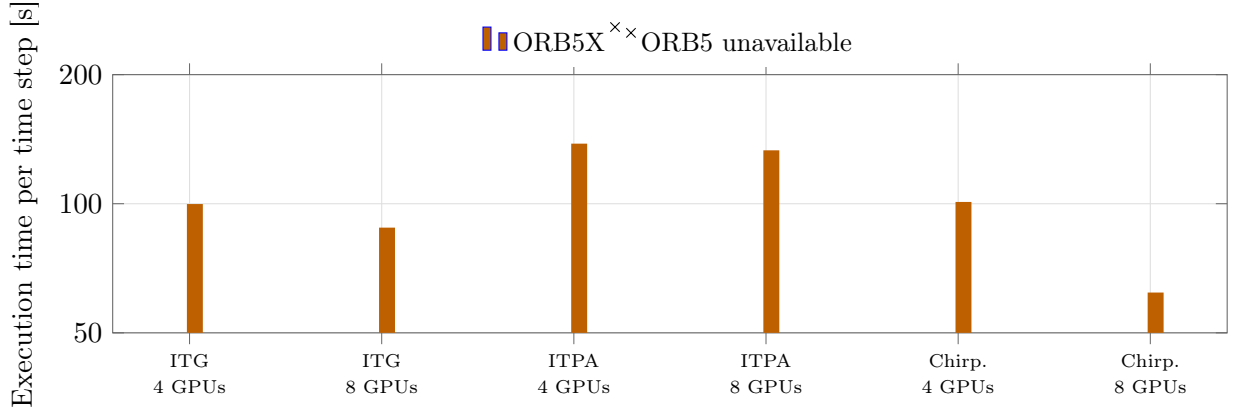

\end{document}